\documentclass[article]{jss}
\usepackage{thumbpdf,lmodern}
\usepackage[utf8]{inputenc}
\usepackage[margin=1.25in]{geometry}
\usepackage{amsmath}
\usepackage{amssymb}
\usepackage{graphicx}
\usepackage{dsfont}
\usepackage{tikz}
\usepackage{algorithm}
\usepackage{algorithmic}
\usepackage{eqparbox}
\usepackage{hyperref}
\graphicspath{ {./images/} }

\makeatletter
\newcommand{\distas}[1]{\mathbin{\overset{#1}{\kern\z@\sim}}}%
\newsavebox{\mybox}\newsavebox{\mysim}
\newcommand{\distras}[1]{%
  \savebox{\mybox}{\hbox{\kern3pt$\scriptstyle#1$\kern3pt}}%
  \savebox{\mysim}{\hbox{$\sim$}}%
  \mathbin{\overset{#1}{\kern\z@\resizebox{\wd\mybox}{\ht\mysim}{$\sim$}}}%
}
\makeatother

\author{Eric Weine\\Department of Statistics\\University of Chicago\\Chicago, IL USA
   \And Mary Sara McPeek\\Department of Statistics\\University of Chicago\\Chicago, IL USA
   \And Mark Abney\\Dept.\ of Human Genetics\\University of Chicago\\Chicago, IL USA}
\Plainauthor{Eric Weine, Mary Sara McPeek, Mark Abney}

\title{Application of Equal Local Levels to Improve Q-Q Plot Testing Bands with \proglang{R} Package \pkg{qqconf}}
\Plaintitle{Application of Equal Local Levels to Improve Q-Q Plot Testing Bands with R Package qqconf}
\Shorttitle{Equal Local Levels Q-Q Plot Testing Bands with \pkg{qqconf}}
\Abstract{
  Quantile-Quantile (Q-Q) plots are often difficult to interpret because it is unclear
how large the deviation from the theoretical distribution must be to indicate a lack of fit.  Most
Q-Q plots could benefit from the
addition of meaningful global testing bands, but the use of such bands unfortunately
remains rare because of the
drawbacks
of current approaches and packages.
These drawbacks include
incorrect global Type I error rate, lack of
power to detect deviations in the tails of the distribution, relatively slow computation for large data sets, and
limited applicability. To solve these problems, we apply the
equal local levels global testing method, which we have implemented in
the \proglang{R}
Package \pkg{qqconf}, a versatile tool to create Q-Q plots and probability-probability (P-P) plots in a wide variety of settings, with
simultaneous testing bands rapidly created using recently-developed algorithms.
\pkg{qqconf} can easily be used to add global testing bands to Q-Q plots made by other packages.  In addition to being quick to compute,
these bands have a variety of
desirable properties, including accurate global levels,
equal sensitivity to deviations in all parts of the null distribution (including the tails), and applicability
to a range of null distributions.  We illustrate the use of \pkg{qqconf} in several applications: assessing normality of residuals from regression, assessing accuracy of $p$~values, and use of Q-Q plots in genome-wide association studies.
}


\Address{
  Eric Weine\\
  Department of Statistics\\
  University of Chicago\\
  5747 S Ellis Ave, Chicago, IL\\
  E-mail: \email{ericweine15@gmail.com}\\
  \\
  Mary Sara McPeek\\
  Department of Statistics\\
  University of Chicago\\
  5747 S Ellis Ave, Chicago, IL\\
  E-mail: \email{mcpeek@uchicago.edu}\\
  URL: \url{https://www.stat.uchicago.edu/~mcpeek/}\\
  \\
  Mark Abney\\
  Department of Human Genetics\\
  University of Chicago\\
  920 E 58th St., Chicago, IL\\
  E-mail: \email{abney@uchicago.edu}\\
  URL: \url{https://hgen.uchicago.edu/program/faculty/mark-abney}
}

\begin{document}

\maketitle

\section{Introduction}
Quantile-Quantile (Q-Q) plots \citep{wilk_gnanadesikan_1968} are a common statistical tool
used for judging
whether
a sample comes from a specified distribution, and,
perhaps most usefully, for visualizing
the particular ways in which the sample might seem to deviate from that distribution. Despite
their ubiquity, they are often difficult to interpret because it is challenging to
determine if the extent of the observed deviation from the specified distribution is
sufficient to indicate a lack of fit as opposed to just being due to sampling variability.
To aid in interpretation, it is useful to put goodness-of-fit testing bands on Q-Q plots.
\newline
\newline
A few methods have been created toward this end (reviewed by
\cite{aldor2013power}). Naively, one could
use a pointwise testing
band, an approach that is equivalent to conducting
 a level-$\alpha$ test on each order statistic of the sample.  However, because of the large
number of tests, the probability that the data ever leave the band is far higher than $\alpha$,
so the pointwise approach does little to help with the problem of interpretability.
To appropriately deal with this
multiple testing problem, the Kolmogorov-Smirnov (KS)
statistic \citep{kolmogoroff1941confidence, smirnov1944approximate}
is sometimes used to create
a simultaneous testing band for a Q-Q plot. While
this
method controls Type I error, the KS test suffers from very low power under a variety of
reasonable alternatives because it has low sensitivity to deviation in the tails of
the null distribution \citep{aldor2013power, berk1979goodness, mason1983}. \\

To overcome this problem, one can instead
apply the equal local levels (ELL) global testing method to create
simultaneous testing bands for Q-Q plots.  The ELL global testing method was
originally introduced by \citet{berk1979goodness} (their $M^+_n$ is a one-sided ELL test
statistic) and further developed by Gontscharuk and colleagues
\citep{gontscharuk2015intermediates, gontscharuk2016goodness,
gontscharuk2017asymptotics} as an improvement over
the higher criticism \citep{donohojin2015} and KS global testing methods.
To conduct the ELL global test at level $\alpha$, one
conducts a ``local'' (or pointwise) test at level $\eta$ on each order statistic of the
sample and rejects the global test whenever at least one of the local
tests is rejected, where the local level
$\eta$ must be chosen so that the global level of the test is the desired value
$\alpha$. The fact that the same local level is applied to each order statistic means
that the ELL testing band can be viewed
as impartial in its sensitivity to deviations from different parts of the null distribution,
a sensible choice for use in a generic tool such as a Q-Q plot.
In the specific context of assessing normality with a Q-Q plot,
\citet{aldor2013power} proposed to apply ELL to create two-sided testing bands by using
simulation to determine the value of $\eta$ needed in each case, a method they called ``tail
sensitive'' (TS) because it has more sensitivity in the tails than KS.  Through a series of
examples and simulations, they effectively demonstrate the superiority of
ELL testing bands over KS for detecting deviations from normality in a Q-Q plot in
a variety of cases of interest.  An
advantage of the
simulation-based approach to computing ELL bands is that it gives a straightforward way to
incorporate the effects of parameter estimation.  However,
such an approach is arguably too slow to be conveniently applied to large, modern datasets.
Considering that the Q-Q plot is meant to be a handy visualization tool and
not an end-goal of analysis, it is important that the bands be virtually instantaneous
to compute on a laptop or they are unlikely to be widely used.  In our ELL implementation,
instead of simulation,
we use pre-computation based on fast algorithms,
supplemented with asymptotic approximations for sample sizes over 100K.\\

Until now, available software for putting testing bands on Q-Q plots has been limited.
The base-\proglang{R} package \pkg{stats} provides functionality for creating a Q-Q plot to compare a sample against the normal distribution, and with a bit more difficulty, one
can create Q-Q plots for other distributions, but it does not provide any way
to put testing bands on those plots. The package \pkg{qqplotr} \citep{qqplotr} provides
a number of helpful additions to the base-\proglang{R} functionality, including the
ability to easily create Q-Q plots for a variety of reference distributions,
the ability to create simultaneous testing bands using KS for a variety of reference
distributions, and the ability to create simultaneous testing bands using TS
only for the normal reference distribution. However,
because it is based on simulation, the TS approach can be a bit slow, taking several
minutes to produce bands for a sample size in the tens of thousands.\\

Our development of the \proglang{R} package \pkg{qqconf} is motivated by two major unmet needs in obtaining testing
bands for Q-Q plots: (1) the need for ELL
testing bands for non-normal distributions, particularly the uniform distribution, and (2) the practical need for greater speed in obtaining ELL
testing bands for Q-Q plots in all cases, including normal.  Regarding (1),
in addition to testing for normality, important uses of
Q-Q plots include assessing accuracy of $p$~values (Section \hyperref[sec:accuracy]{3.2}) and
applications in genomics (Section \hyperref[sec:gwas]{3.3}) both of which involve assessing
uniformity, so it would be extremely useful to have ELL simultaneous testing bands for
Q-Q plots for the uniform case in particular, as well as
for other non-normal distributions in
general.  Regarding (2), in light of the demonstrated superiority of ELL over other
approaches for creating testing bands for Q-Q plots, one of our major software goals is to
make creation of ELL testing bands (at least for $\alpha =$ .05 or .01) so fast that this approach
can confidently be used as the default for all Q-Q plots, without concern for
taxing the casual user's
patience or processing resources.\\

In what follows, we introduce the \proglang{R} package \pkg{qqconf} (available on CRAN) for making Q-Q and
probability-probability (P-P) plots.  The \code{get_qq_band} 
function in \pkg{qqconf} can quickly provide ELL testing
bands for comparing even very large samples to any reference distribution with a quantile
function (e.g., \code{qnorm}, \code{qchisq})
implemented. In addition to these testing bands, which can be output for use with other
plotting packages, \pkg{qqconf} provides a variety of
plotting functionalities that allow the user to easily visualize where any
deviation of the sample from the null distribution may occur.  In Section
\hyperref[sec:methods]{2}, we introduce the methods required for computation of ELL
testing bands for Q-Q plots.
In Section \hyperref[sec:examples]{3}, we demonstrate the functionality of \pkg{qqconf}
in applications including assessing normality of residuals from regression
(Section \hyperref[sec:regression]{3.1}), assessing accuracy of $p$~values (Section
\hyperref[sec:accuracy]{3.2}), and use of Q-Q plots in genome-wide association studies
(Section \hyperref[sec:gwas]{3.3}).\\

\section{Methods}
\label{sec:methods}

\subsection{Local levels for global hypothesis testing}

Our ELL method for creating appropriate simultaneous testing bands for Q-Q plots can be viewed as an application of
the following more general testing framework.
Suppose we have real-valued observations
\begin{equation*}
  X_{1}, ..., X_{n} \distras{iid} F,
\end{equation*}
with order statistics $X_{(1)}\leq X_{(2)} \leq \ldots \leq X_{(n)}$,
and we are interested in conducting the following hypothesis test at level $\alpha$:
\begin{equation*}
      H_{0}: F = F_{0} \textrm{ vs. } H_{A}: F \neq F_{0},
\end{equation*}
where we refer to $H_0$ as the ``global null hypothesis'' and $\alpha$ as the ``global level'', and where
$F_0$ is a known continuous distribution on $\mathbb{R}^1$ (or on some finite or infinite sub-interval
of $\mathbb{R}^1$ such as (-1,1) or (0,$\infty$)).
For simplicity, we start by assuming that all parameters of $F_0$ are known (we relax this
assumption later).
One approach to this hypothesis testing problem, referred to as ``local levels'' \citep{gontscharuk2016goodness}, is
to conduct $n$ separate (``local'') hypothesis tests, one on each of the order statistics $X_{(1)}, \ldots, X_{(n)}$,
where the test on the
$i$th order statistic has level $\eta_{i}$ (the $i$th local level). Then, one rejects the global null hypothesis if at
least one of the $n$ local
tests results in a rejection.
That is, we construct a set of intervals
\begin{equation*}
    (h_{1}, g_{1}), ..., (h_{n}, g_{n}),
\end{equation*}
where $h_i < g_i$ for $1\leq i \leq n$, and under the null hypothesis, P($X_{(i)} \notin (h_i,g_i)) = \eta_i$,
and we reject $H_{0}$ if
\begin{equation}
    X_{(i)} \not\in (h_{i}, g_{i}) \mbox{ for at least one value of $i$ such that } 1 \leq i \leq n.\label{rejectionevent}
\end{equation}
In this general setting, the level $\alpha$ of the global test is determined by the
vectors of lower and upper interval endpoints, $(h_1,\ldots, h_n)$ and $(g_1,\ldots, g_n)$
and the null cdf $F_0$.

\subsection{ Two-sided ELL}

For the Q-Q plot application, we want to create level-$\alpha$ testing bands that are ``agnostic''
to any alternative distribution. By this, we mean that we would like to design a local levels test
such that, firstly, the global test applies equal scrutiny to each order statistic, i.e., we set the
local levels to be equal:
\begin{equation}
    \eta_{1} = \eta_{2} = ... = \eta_{n} = \eta,\label{equallevels}
\end{equation}

and, secondly, the local tests give equal weight to deviations of $F$ from $F_0$ in either direction, i.e.,
we choose
\begin{equation}
    h_{i} = F^{-1}_{0i}(\eta / 2) \text{ and } g_{i} = F^{-1}_{0i}(1 - \eta / 2),\label{equalsides}
\end{equation}
where $F_{0i}$ is the cdf of the $i$th order statistic under the null hypothesis, which is easily
obtained from $F_0$ (see, e.g., Section 5.4 of \citep{casellaberger2002}).   We refer to the global test derived from local levels under conditions
(\ref{equallevels}) and (\ref{equalsides}) as the two-sided ELL test.\\

The main difficulty in applying the two-sided ELL test is in determining the local level $\eta$ that will
result in the desired global level $\alpha$.  One nice property of the two-sided ELL is that the
local level $\eta$ needed to achieve global level $\alpha$ depends only on $\alpha$ and on the sample
size $n$, and not on $F_0$ at all.  This can be seen by noting that under the null hypothesis,
\begin{equation}
    F_{0}(X_{1}), ..., F_{0}(X_{n}) \distras{iid} U(0, 1).\label{wolog}
\end{equation}
Thus, without loss of generality, we can take the null distribution to be $U(0,1)$ and determine
the needed $\eta$ and the interval endpoints $(h_1,\ldots, h_n)$ and $(g_1,\ldots, g_n)$ for this
case.  To convert back to the original scale, all that is needed is to apply $F_0^{-1}$ to each
of the resulting interval endpoints.

\subsection{Calculation of the local level for two-sided ELL}
\label{sec:alg}

Given the sample size $n$ and the desired global level $0 < \alpha < 1$, we define $\eta_n(\alpha)$ to be the
local level $\eta$ that will result in global level $\alpha$ for the ELL test.
Note that $\eta_n(\alpha)$ is a continuous, monotone increasing function of $\alpha$, and we denote its
inverse by $\alpha_n(\eta)$.  Given $n$ and $\alpha$, the basic approach to obtaining $\eta_n(\alpha)$
involves a binary search over $\eta \in (0,1)$, where for each
value of $\eta$, we obtain $(h_1,\ldots,h_n)$ and $(g_1,\ldots,g_n)$ via Equation
(\ref{equalsides}), and then we calculate $\alpha_n(\eta)$, the probability of the event described in
Equation (\ref{rejectionevent}), i.e., we find that probability that $(X_{(1)}, \ldots,
X_{(n)})$ falls outside the region $(h_1,g_1)\times\cdots\times (h_n,g_n)$.
Then we perform a binary search to find the $\eta$ such that
$\alpha_n(\eta)= \alpha$, the desired global level.
\newline
\newline
To calculate $\alpha_n(\eta)$, several recursive approaches have previously been
developed (see \cite{shorackwellner2009}), as well as a fast Fourier
transform (FFT) based approach \citep{moscovichnadler2016}.
In \pkg{qqconf}, we apply the method of \cite{moscovichnadler2016}, as implemented in \cite{crossprob}, which
can be used by the ELL method to obtain simultaneous Q-Q plot testing bands at global level $\alpha$ for
any $n$ and $\alpha$.  In addition, \pkg{qqconf} offers a faster approximate approach specifically
for the most commonly-used global levels of $\alpha =.05$ and .01.  To do this we have applied
our own recursive formula (Appendix \hyperref[sec:appa]{A}) for obtaining $\alpha_n(\eta)$ in order
to generate look-up tables for $\eta_n(\alpha)$ for $\alpha = .05$ and .01 with
sample sizes $n$ up to 1 million and 500K, respectively, where the tables are
relatively dense for $n$ up to 100K.
If the user inputs $\alpha = .05$ or .01 with a value of $n$ less than or equal to 100K, we either return back the pre-computed value of $\eta$ if $n$ happens to
be a grid point, or we use linear interpolation if the value of $n$ is between grid points, which leads to a highly accurate
approximation. If the user inputs a value of $n$ greater than 100K with $\alpha = .05$ or .01, we use the asymptotic
approximation given in Section \hyperref[sec:approx]{2.4}.
This allows \pkg{qqconf} to provide essentially instantaneous simultaneous testing bands for the cases
$\alpha =.05$ and .01 for any reference distribution with quantile function implemented,
with the FFT approach \citep{moscovichnadler2016}
used primarily for fast ``on-the-fly'' calculations with other choices of $\alpha$.

\subsection{Local level approximations in large samples}
\label{sec:approx}
For sufficiently large values of the sample size $n$ (or, equivalently, the number of local tests),
it can be expedient to apply an accurate asymptotic approximation of
$\eta_n(\alpha)$ in place of exact computation. Previous authors \citep{gontscharuk2017asymptotics}
showed that
an asymptotic approximation of
$\eta_n(\alpha)$ is
\begin{equation*}
    \eta_{asymp} = \frac{-\log(1 - \alpha)}{2\log(\log(n))\log(n)}.
\end{equation*}
However, as they note, this approximation gives poor performance for $n$ even as large as $10^{4}$. To improve this approximation, they propose to add a smaller order correction term, resulting in an approximation of the form
\begin{equation}
    \eta_{approx} = \frac{-\log(1 - \alpha)}{2\log(\log(n))\log(n)}\left[1 - c_{\alpha}\frac{\log(\log(\log(n)))}{\log(\log(n))}\right], \label{approxform}
\end{equation}
where $c_{\alpha}$ is chosen empirically. For the values $\alpha = .01, .05, \textrm{ and } .1$ they chose $c_{\alpha} = 1.6, 1.3,
\textrm{ and } 1.1$, respectively. To select these $c_{\alpha}$ values,
the authors calculated the values of $\eta_n(\alpha)$ to high precision on
a grid of values up to $n = 10,000$. \\

We performed more extensive tests of these approximations for the cases $\alpha = .01$ and .05.  To do
this, we calculated the values of $\eta_n(\alpha)$
with high precision on a grid of values up to $n = 500,000$ for $\alpha = .01$ and up to
$n = 10^6$ for $\alpha = .05$. Based on our evaluation, we find that $c_{\alpha} = 1.3$ is
satisfactory for $\alpha = .05$, but that $c_{\alpha} = 1.6$ for $\alpha = .01$ is not sufficiently
accurate for our purposes.
We instead found that $c_{\alpha} = 1.591$ led to better performance for $\alpha = .01$.  For example, for
$n$ in the range of 15K to 500K, the absolute
relative error in the approximation based on $c_{\alpha} = 1.6$ is
always more than $.0067$, while that based on $c_{\alpha} = 1.591$ is always less than .001.
\newline
\newline
We implement these asymptotic approximations in \pkg{qqconf} as part of our faster approximate approach
specifically for $\alpha = .01 \textrm{ and } .05$ with $n>$ 100K,
as described in Section \hyperref[sec:alg]{2.3}.
(Our package also implements the approximation given in Equation (\ref{approxform}) for $\alpha = .1$ with $c_{\alpha} = 1.1$.)

\subsection{One-sided ELL}
\label{sec:one-sided}

In some instances, a one-sided version of ELL is of particular interest.  For example, suppose $X_1,
\ldots, X_n$ are $p$~values, with $X_i$ representing the $p$~value of the $i$th hypothesis test, which has
corresponding null
hypothesis $H_0^{(i)}$, where
$X_1,\ldots, X_n$ are assumed to be independent, with $X_i \sim$ U(0,1) if $H_0^{(i)}$ is true.  Suppose
we are interested in testing the global null
hypothesis $H_0:$ all of $H_0^{(1)}, \ldots H_0^{(n)}$ are true against the alternative $H_A:$ at least one
of $H_0^{(1)}, \ldots H_0^{(n)}$ is false.  Within the equal local levels framework, we would typically
do this by assuming
\begin{equation*}
  X_{1}, ..., X_{n} \distras{iid} F,
\end{equation*}
and testing the null hypothesis $H_0: F(x) = x$ for
all $x \in (0,1)$ vs.\ the one-sided alternative $H_A: F(x)>x$ for
at least one $x \in (0,1)$.
In this case, a one-sided test is commonly used because one is typically only interested in
$p$~values that are smaller than expected, not larger than expected.
This is exactly the context considered in
\cite{berk1979goodness}, in which the ideas behind ELL were first laid out.\\

More generally, one could test
\[ H_0: F = F_0 \mbox{ for all } x \in \mathbb{R} \mbox{ vs. } H_A: F > F_0 \mbox{ for some }
x \in \mathbb{R}.\]
In this context, a one-sided global test of $H_0$ based on local levels $\eta_1,
\ldots, \eta_n$ would involve first constructing a set of lower bounds $h_1,\ldots,h_n$, where
\begin{equation}
h_i = F_{0i}^{-1}(\eta_i),\label{equalsides_onesided}
\end{equation}
and then rejecting if
\begin{equation*}
 X_{(i)} < h_i \mbox{ for at least one value of } i \mbox{ such that } 1 \leq i \leq n.\label{rejectionevent_onesided}
\end{equation*}
We define
the one-sided ELL test with global level $\alpha$ to be the test of this type obtained by setting $\eta_1 =
\cdots = \eta_n = \eta$ and choosing $\eta$ to obtain global level $\alpha$.\\

Given the sample size $n$ and the desired global level $0 < \alpha < 1$, we define
$\eta^{'}_n(\alpha)$ to be the
local level $\eta$ that will result in global level $\alpha$ for the one-sided ELL test.
As in the two-sided case, we denote the inverse function of $\eta^{'}_n(\alpha)$ by $\alpha^{'}_n(\eta)$.
Given $n$ and $\alpha$, we obtain $\eta^{'}_n(\alpha)$
by a binary search over $\eta \in (0,1)$, where for each
value of $\eta$, we obtain $(h_1,\ldots,h_n)$ via Equation
(\ref{equalsides_onesided}), and then we calculate $\alpha^{'}_n(\eta)$,
the probability
that $(X_{(1)}, \ldots,
X_{(n)})$ falls outside the region $(h_1,1)\times\cdots\times (h_n,1)$.
Then we perform a binary search to find the $\eta$ such that
$\alpha^{'}_n(\eta)= \alpha$, the desired global level.
\newline
\newline
To calculate $\alpha^{'}_n(\eta)$, several approaches have previously been
developed \citep{shorackwellner2009, moscovich2020}.  \pkg{qqconf} currently uses the method of
\cite{moscovichnadler2016} as implemented in \cite{crossprob}.
We have also implemented two recursive approaches, described in
Appendix \hyperref[sec:appb]{B}: an exact version and
an approximate version that is much faster and
bounds the relative error in the reported global significance level to a tolerance set by the user.

\subsection{Additional implementation issues}

To create the ``expected'' quantiles for a Q-Q plot, we apply the inverse cdf $F_0^{-1}$ to a set
of probability points.  For the normal distribution, it has been shown \citep{blom1958} that the
means of the order statistics of $n$ i.i.d.\ draws are well-approximated by the 
the above process 
when \code{ppoints(n)} is used to generate the probability points, while
for the uniform distribution, the means of the order statistics are obtained exactly 
when \code{ppoints(n,a=0)} is 
used.  For other distributions, appropriate approximations to the means of the order statistics
could be obtained on a case-by-case basis.  
(Because a P-P plot is basically a variation on a uniform Q-Q plot, 
the exact mean probability points for a P-P plot are obtained for all distributions by 
\code{ppoints(n,a=0)}.)  For creating
the ``expected'' line in a Q-Q plot, we
propose the medians of the order statistics as a useful alternative to their means.  Exact medians 
of the order statistics for i.i.d.\ draws from any distribution can easily be obtained
by applying the inverse cdf to \code{qbeta(.5,c(1:n),c(n:1))}.  The resulting 
``expected'' line is the unique line that is guaranteed to lie completely within the ELL
band, regardless of the global level $\alpha$ or the distribution.  All 3 of the above expected 
lines are options within \pkg{qqconf}.\\

The most commonly-encountered uses of Q-Q plots are to assess normality in various contexts and to assess
uniformity of $p$~values for a set of independent hypothesis tests, and we give examples of both in
Section \hyperref[sec:examples]{3}.  When assessing normality, typically the mean $\mu$ and
standard deviation $\sigma$ would not be
known but would need to be estimated from the data in order to make either an
``expected'' line or any kind of testing band for a Q-Q plot.  For example, in \pkg{base}-\proglang{R} the function
\code{qqline} makes an expected line that by default passes through the first and third quartiles, which is equivalent to
estimating $\mu$ by the mid-quartile and $\sigma$ by the inter-quartile range
multiplied by .7413.\\

\begin{table}
\caption{Empirical type 1 error at nominal level .05 for testing normality with different parameter estimation methods, based on $10^4$ simulation replicates.\label{table:type1err}}
\begin{center}
\begin{tabular}{c|ccccc}
& \multicolumn{5}{c}{Empirical type 1 error (se) when using}\\
Sample size & Sample sd & MAD & $Q_n$ & $S_n$ & True\\
\hline
100 & .0011 (.0003) & .0963 (.0030)& .0278 (.0016)& .0427 (.0020) & .0522 (.0022)\\
500 & .0035 (.0006)& .1015 (.0030)& .0260 (.0016)&  .0443 (.0021)& .0514 (.0022)\\
10,000 & .0111 (.0010)& .1040 (.0031)& .0327 (.0018)& .0498 (.0022)& .0480 (.0021)\\
\end{tabular}
\end{center}
\end{table}

In \pkg{qqconf}, the default is to estimate $\mu$ by the
median and $\sigma$ by the estimator $S_n$ of \cite{rousseeuw1993},
where $S_n$ is a highly robust scale estimator with very low gross-error sensitivity that
is more efficient than median-absolute-deviation (MAD) and
approximately unbiased even in small sample sizes.  To validate this choice, we performed
simulation studies under the null hypothesis of normality and assessed the type
1 error of the 5\% rejection bounds generated by ELL, where we used one of 5 choices for $(\mu,\sigma)$: (1) sample mean and sample s.d., (2) sample median and sample MAD,
(3) sample median and $Q_n$, another estimator of $\sigma$ discussed by \cite{rousseeuw1993},
(4) sample median and $S_n$, and (5) the true values of $\mu$ and $\sigma$ for comparison,
and where these are denoted in Table \ref{table:type1err} by ``sample sd'', ``MAD'', ``$Q_n$'', ``$S_n$'' and ``true", respectively.
We note that the entire simulation study is invariant to the choice of true mean and s.d., because these just become location and scale factors for all the data and the estimators and
therefore cancel out in the type 1 error assessment.   The results for n=100,
500 and $10^4$ are given in Table \ref{table:type1err}, where we can see that using median and $S_n$ gives type 1 error very close to the nominal level, though slightly conservative for small sample sizes.
Methods to handle parameter uncertainty with an exact calculation (as opposed to simulation) have
been discussed in the context of the normal distribution \citep{rosenkrantz2000confidence}, but a
general method
towards this end has not been developed.  Use of Q-Q plots for
distributions other than normal for which the parameters are unknown is rarer, and for those
cases the current default in \pkg{qqconf} is maximum likelihood estimation, though the user can replace
that with an estimate of their choice.  Note that in applications such as assessing uniformity of $p$~values
(Section \hyperref[sec:accuracy]{3.2})
or in the genomics example in Section \hyperref[sec:gwas]{3.3}, no parameter estimation is required.

\section{Examples}
\label{sec:examples}
One of the main advantages of the local levels method compared to other global testing approaches is that it can easily be used to put testing bands onto Q-Q plots by simply graphing each $(h_{i}, g_{i})$ interval. This allows us to examine how a dataset might deviate from some null distribution much better than simply applying a test that yields a binary conclusion. Below, we present a few examples where a Q-Q plot is useful, and where the local levels test seems ideal for assessing deviation from a global null hypothesis.

\subsection{Assessing normality of residuals from regression}
\label{sec:regression}
When performing an ordinary least squares (OLS) regression, it is common to assume that the error terms
are independently drawn from a normal distribution, e.g., $Y = X\beta + \epsilon$, where $Y_{n\times 1}$
and $X_{n\times p}$ are observable, $\beta_{p\times 1}$ is an unknown parameter vector, and conditional
on $X$, $\epsilon=(\epsilon_1,\ldots,\epsilon_n)^\top$ is assumed to satisfy
\begin{equation}
  \epsilon_{1}, ..., \epsilon_{n} \distras{iid} N(0, \sigma^{2}).\label{olsassumption}
\end{equation}
After obtaining the OLS estimate $\hat{\beta}$ and the residual vector $r =Y - X\hat{\beta}$,
we would like a Q-Q plot of the residuals with a normal reference distribution to aid in testing assumption
(\ref{olsassumption}) above.  (Strictly speaking, even if assumption (\ref{olsassumption}) is correct, the residuals are not
independent because OLS leads to an estimator $\hat{\boldsymbol\beta}$ such that $\sum_{i=1}^{n}r_{i} = 0$.
However, with a reasonable sample size the resulting slight negative
correlation between residuals becomes negligible.)  Without prior
reason to believe that the errors may deviate from normality at any particular point in the distribution, it makes sense to use ELL bands in this case. This is very easy to do with \pkg{qqconf}, as we show below.
\newline
\newline
First, we generate data to perform a regression. Here, we generate each $\epsilon_{i}$ independently from a $t(3)$ distribution.

\begin{Schunk}
\begin{Sinput}
R> set.seed(20)
R> n <- 100
R> x <- runif(n)
R> eta <- rt(n, df = 3)
R> y <- x + eta
\end{Sinput}
\end{Schunk}

Then, we fit a regression with the simulated data

\begin{Schunk}
\begin{Sinput}
R> reg <- lm(y ~ x)
\end{Sinput}
\end{Schunk}

\begin{figure}
\begin{center}
\includegraphics[width=.67\linewidth]{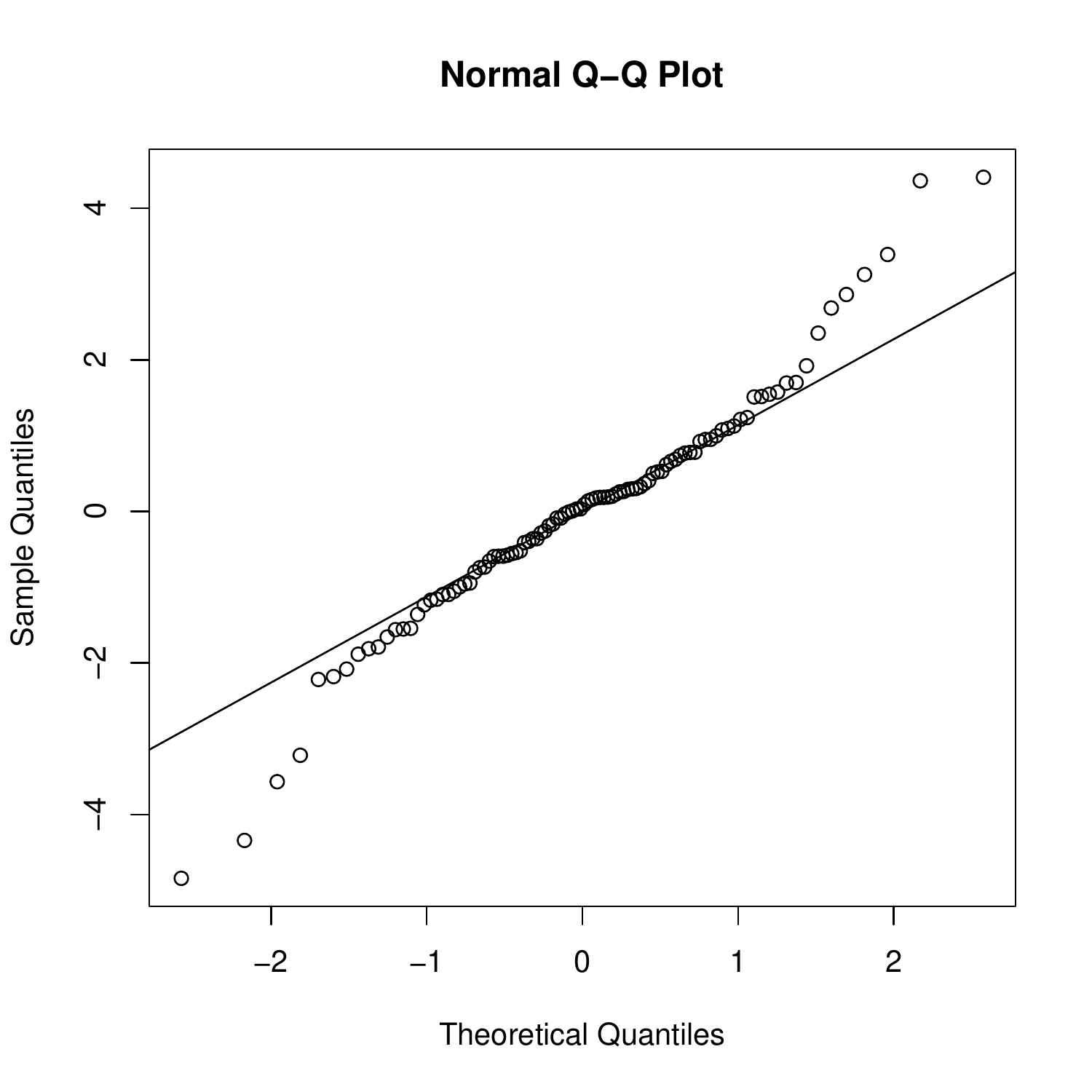}
\end{center}
\caption{Q-Q plot for regression residuals with \pkg{base}-\proglang{R} functionality.}
\label{fig:baserqq}
\end{figure}

Figure \ref{fig:baserqq} shows a Q-Q plot created with \pkg{base}-\proglang{R} functionality
using the function \code{qqnorm} and \code{qqline}, as follows:

\begin{Schunk}
\begin{Sinput}
R> qnorm_plot <- qqnorm(reg$residuals)
R> qqline(reg$residuals)
\end{Sinput}
\end{Schunk}

\begin{figure}
\begin{center}
\includegraphics[width=.67\linewidth]{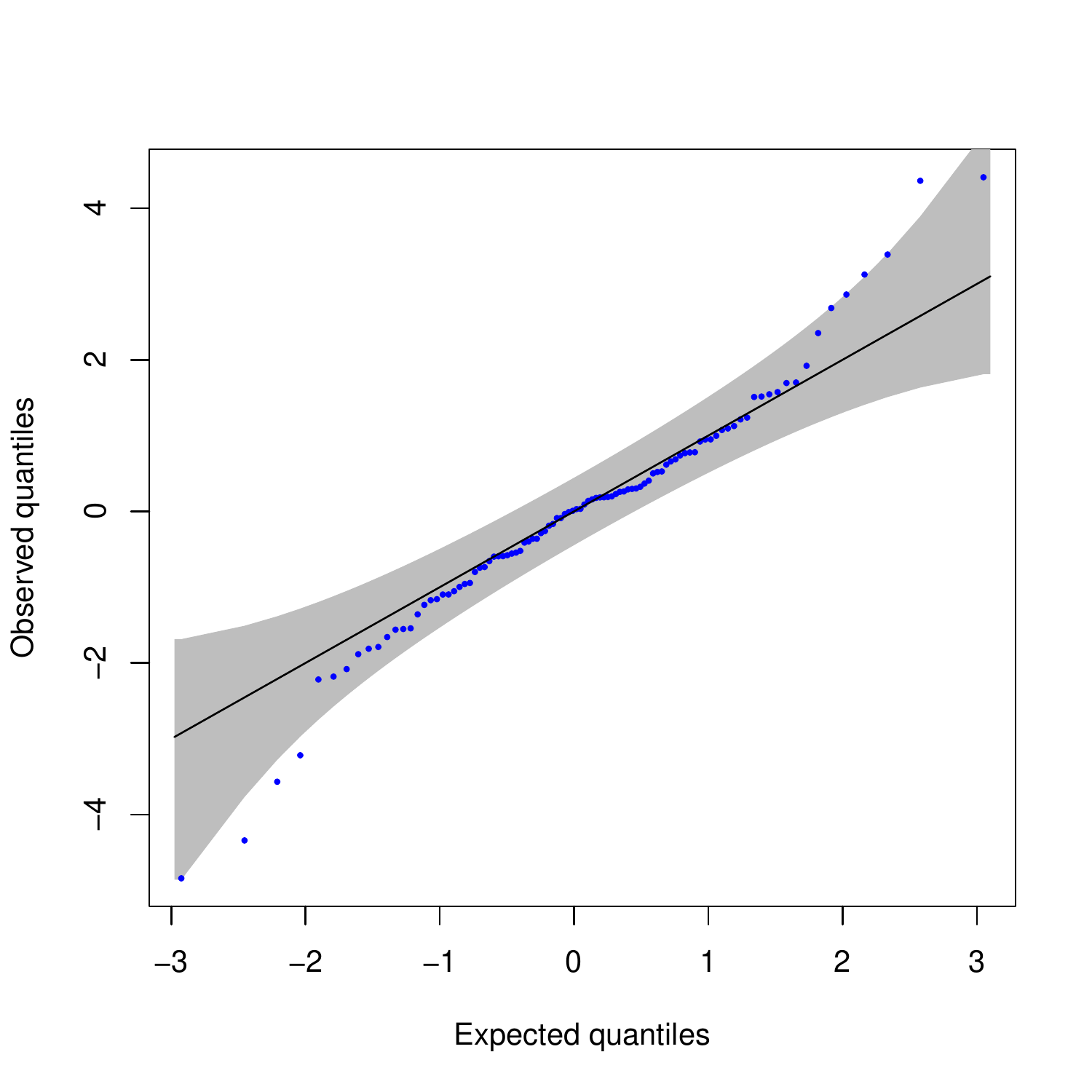}
\end{center}
\caption{Q-Q plot for regression residuals with ELL bounds using \pkg{qqconf}.}
\label{fig:ex1ellqq}
\end{figure}

Clearly there is some indication of deviation from normality in Figure \ref{fig:baserqq},
but it can be hard to tell how significant the deviation is without a testing band.
In Figure \ref{fig:ex1ellqq}, we improve upon this by using \code{qq_conf_plot}
to create a Q-Q plot with a .05-level ELL testing band as follows:

\begin{Schunk}
\begin{Sinput}
R> qqconf::qq_conf_plot(
+    obs = reg$residuals,
+    points_params = list(col = "blue", pch = 20, cex = .5)
+  )
\end{Sinput}
\end{Schunk}

In Figure \ref{fig:ex1ellqq}, we can clearly see that both the left and right tails of the residuals go
beyond the normal testing bounds, giving strong evidence that the errors were not generated from a normal distribution.\\

\begin{figure}
\begin{center}
\includegraphics[width=.67\linewidth]{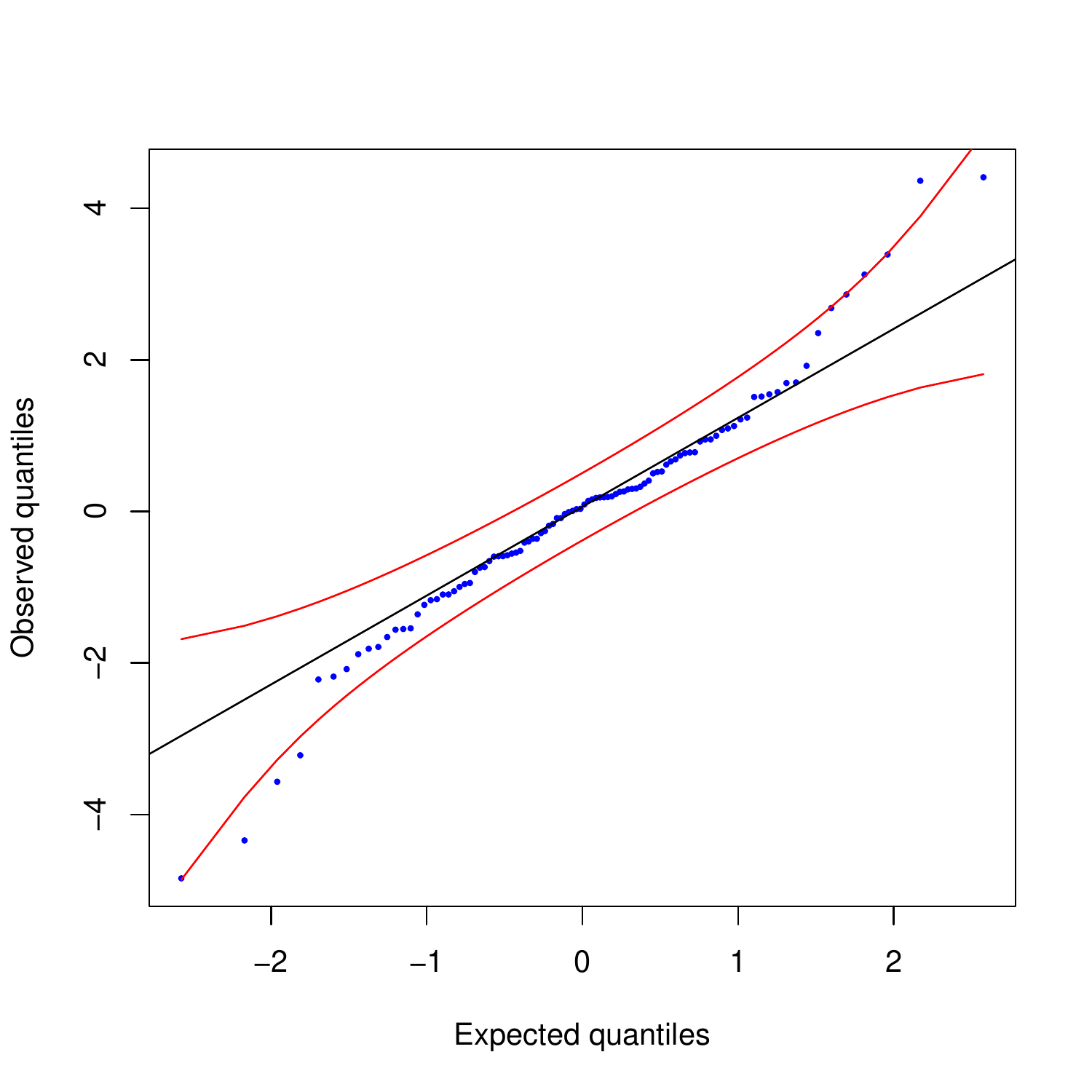}
\end{center}
\caption{Q-Q plot for regression residuals with ELL bands from \pkg{qqconf} added to a plot made with \pkg{base}-\proglang{R} \code{qqnorm}.}
\label{fig:ex1baserbands}
\end{figure}

If the user prefers to use another plotting software, \pkg{qqconf} also provides a separate interface, \code{get_qq_band}, for obtaining the testing band iself. 
The \code{band_method} argument of \code{get_qq_band} 
allows for ELL, KS or pointwise bands to be created.
The band computed by \code{get_qq_band} can easily be used with, e.g., 
base-\proglang{R}'s \code{qqnorm} as below, producing Figure \ref{fig:ex1baserbands}.  When adding
a testing band to a Q-Q plot produced outside of \pkg{qqconf}, in order to have the correct
type 1 error for the band, it is essential that the same
x-coordinates be used for plotting both the data points and the bound points for the band.  This
is accomplished in Figure \ref{fig:ex1baserbands} by use of \code{sort(qnorm_plot$x)} as the
x-coordinates for the upper and lower bounds for the band in the code below, as these were the
x-coordinates used by \code{qnorm} to plot the data points.
Figure 3 is generated as follows:

\begin{Schunk}
\begin{Sinput}
R> band <- qqconf::get_qq_band(obs = reg$residuals)
R> plot(
+    qnorm_plot,
+    col = "blue",
+    pch = 20,
+    cex = .5,
+    xlab = "Expected quantiles",
+    ylab = "Observed quantiles"
+  )
R> lines(sort(qnorm_plot$x), band$lower_bound, col = "red")
R> lines(sort(qnorm_plot$x), band$upper_bound, col = "red")
R> qqline(
+    qnorm_plot$x,
+    datax = TRUE,
+    distribution = function(p) qnorm(
+      p, mean = band$dparams$mean, sd = band$dparams$sd
+    )
+  )
\end{Sinput}
\end{Schunk}

Moreover, if the user prefers to use \pkg{qqplotr}, this can also be done easily, as shown in 
Figure \ref{fig:ex1gg}.  Again, it is critical that the same x-coordinates be used to 
plot both the 
points and the bounds of the band. This is accomplished in the code below for 
Figure \ref{fig:ex1gg} by setting 
\code{band_df$expected <- build_plot$data[[1]]$x}, which puts the x-coordinates that will be
used to plot the points
into \code{band_df$expected}, and then using \code{x = expected} as an argument to \code{aes} in
the call to \code{geom_ribbon} that creates the testing band.  Figure \ref{fig:ex1gg} is generated
as follows:

\begin{figure}
\begin{center}
\includegraphics[width=.67\linewidth]{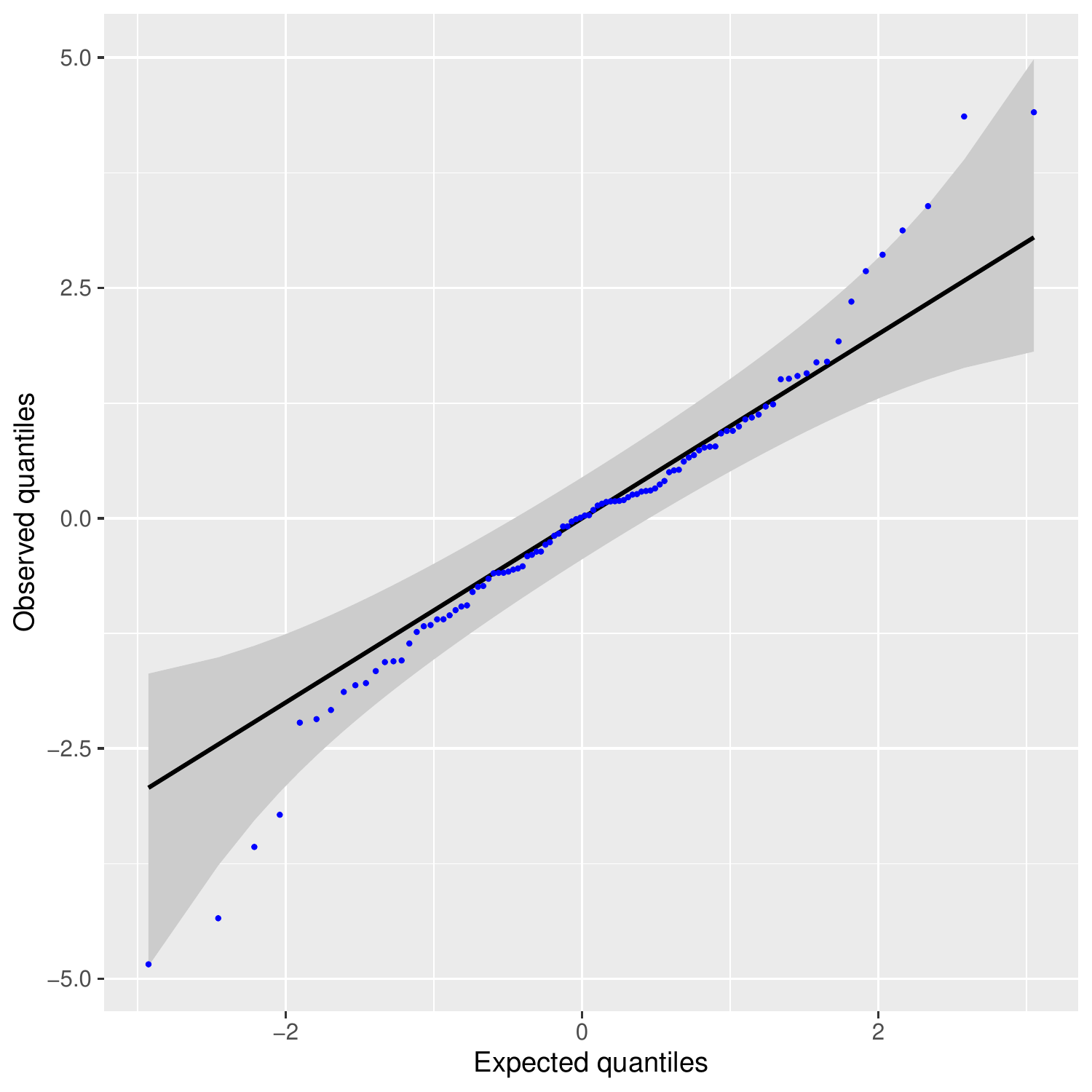}
\end{center}
\caption{Q-Q plot for regression residuals with with ELL bands from \pkg{qqconf} added to a plot made with \code{qqplot}.}
\label{fig:ex1gg}
\end{figure}

\begin{Schunk}
\begin{Sinput}
R> band_df <- data.frame(
+    lower = band$lower_bound,
+    upper = band$upper_bound,
+    obs = reg$residuals
+  )
R> build_plot <- ggplot2::ggplot_build(
+    ggplot2::ggplot(data = band_df, mapping = ggplot2::aes(sample = obs)) +
+      qqplotr::stat_qq_point(dparams = band$dparams)
+  )
R> band_df$expected <- build_plot$data[[1]]$x
R> ggplot2::ggplot(data = band_df, mapping = ggplot2::aes(sample = obs)) +
+    ggplot2::geom_ribbon(
+      ggplot2::aes(ymin = lower, ymax = upper, x = expected),
+      fill = "grey80"
+    ) +
+    qqplotr::stat_qq_line(dparams = band$dparams, identity = TRUE) +
+    qqplotr::stat_qq_point(dparams = band$dparams, color = "blue", size = .5) +
+    ggplot2::xlab("Expected quantiles") +
+    ggplot2::ylab("Observed quantiles")
\end{Sinput}
\end{Schunk}

This example also highlights the advantages of the ELL method over KS. Because KS is much more sensitive
to deviations in the center of the distribution than it is to deviations in the tails of the
distribution, it does not yield a rejection of the null hypothesis in this case (Figure \ref{fig:ex1ksqq}). We generate the KS bounds in Figure \ref{fig:ex1ksqq} by simply setting the \code{method} argument to \code{"ks"} in the \code{qq_conf_plot} function as follows:

\begin{figure}
\begin{center}
\includegraphics[width=.67\linewidth]{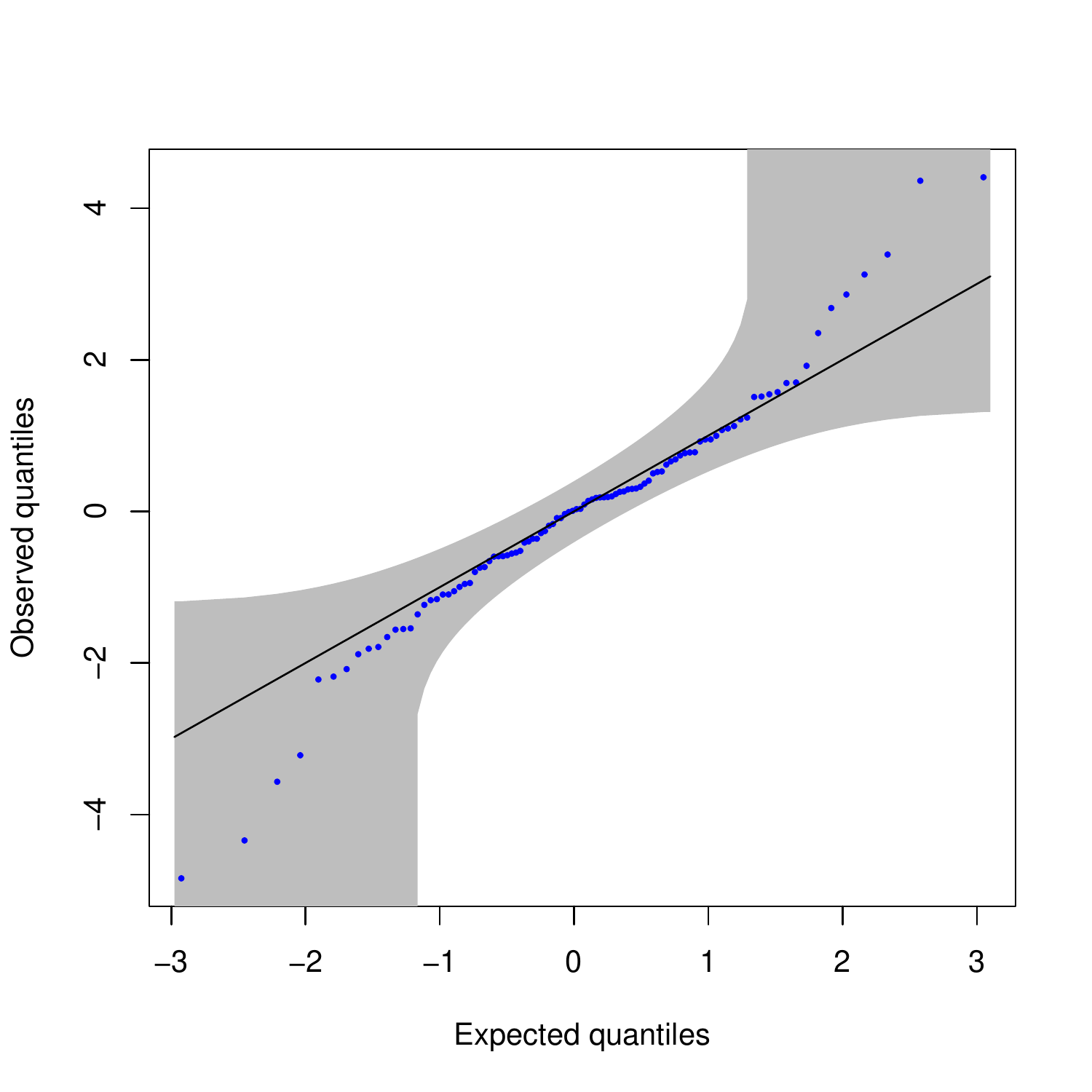}
\end{center}
\caption{Q-Q Plot for regression residuals with KS bounds.}
\label{fig:ex1ksqq}
\end{figure}

\begin{Schunk}
\begin{Sinput}
R> qqconf::qq_conf_plot(
+    obs = reg$residuals,
+    method = "ks",
+    points_params = list(col = "blue", pch = 20, cex = .5)
+  )
\end{Sinput}
\end{Schunk}

\subsection{Q-Q plots for assessing accuracy of $p$ values}
\label{sec:accuracy}

Suppose we have devised a new testing procedure to test a null hypothesis $H_{0}$ with test statistic $T$, where we also specify a particular method to calculate or approximate $p$~values. In such a situation it is important to perform some simulations under the null hypothesis and check that the resulting $p$~value distribution is approximately uniform in the simulation experiment.\\

Typically, the verification of Type I error rate is done using the following procedure:\\

(1) Generate $n$ simulated datasets under $H_{0}$, and calculate $T$ for each simulated dataset to obtain $T_{1}, ..., T_{n}$.
\newline
\newline
(2) Select a value of $\alpha$, and for each of $T_{1}, ..., T_{n}$, determine whether the null hypothesis is rejected at level $\alpha$. Let $N_{\alpha}$ be the observed number of the $n$ tests that are rejected at level $\alpha$.
\newline
\newline
(3) Let $\alpha^{*}$ denote the true probability of rejection under the above procedure. Test the null hypothesis $H_{0}: \alpha^{*} = \alpha$ by applying, e.g., a Z-test of proportions or an exact binomial test to the data $N_{\alpha}$.\\

While the above procedure provides reliable information about the Type I error calibration for one level of $\alpha$, it provides little information about the global calibration of $p$~values. To obtain a useful visualization of the overall performance of the $p$~value calculation method, we instead suggest the following procedure:
\newline
\newline
(1) As above.
\newline
\newline
(2) For each $T_{i}$, calculate the corresponding $p$~value, $p_{i}$, to obtain $p_{1}, ..., p_{n}$.
\newline
\newline
(3) Make a Q-Q plot comparing $p_{1}, ..., p_{n}$ to a $U(0, 1)$ distribution, and apply the local levels procedure to create a simultaneous testing band for the null hypothesis that $p_{1}, ..., p_{n} \distras{iid} U(0, 1)$.
\newline
\newline
This allows us to easily visualize the global calibration of the $p$~values with just one graph and diagnose any issues if they exist. In Step 3, one could use many different testing bands. However, in the calibration of $p$~values, we typically don't have the expectation that our $p$~values would be more likely to deviate from uniform in any particular region, and so it makes sense to use the local levels test because it is agnostic to the space of alternative distribution. Moreover, since it is generally most concerning if small $p$~values are not calibrated (i.e., those in the lower tail of the uniform distribution), the local levels test is preferable to the standard KS test because it is much more sensitive in the tails \citep{aldor2013power}.

\subsubsection{Chi-square test for independence in a $2$ X $2$ table}
We apply this approach to assess the calibration of $p$~values from the Pearson Chi-Square test for independence in a $2 \times 2$ table.  A well-known rule of thumb is that the chi-square test is appropriate as long as the expected cell count in each cell under the null hypothesis is at least 5.  We fix the cell probabilities under the null hypothesis and consider two different cases: in scenario 1, the sample size is $s=200$ and the rule of thumb holds, and scenario 2, the sample size is only $s=20$ and the rule of thumb does not hold.  We use the local levels approach to generate simultaneous testing bands to assess the calibration of the $p$~values from the Pearson Chi-Square test for these two scenarios.\\

More specifically, in each scenario, we randomly generate $n=1000$ $2 \times 2$ tables under the null hypothesis, where each table contains $s$ observations, with $s=200$ in scenario 1 and $s=20$ in scenario 2.  For each table, the $s$ observations are i.i.d.\ with probability $q_{i,j}$ of falling in cell $(i,j)$, for $i=0,1, j=0,1$, where $q_{1,1} = a*b$, $q_{1,0} = a*(1-b)$, $q_{0,1} = (1-a)*b$, and $q_{0,0} = (1-a)*(1-b)$, with a=.15 and b=.4.  For each table, let $X_{i,j}$ denote the observed count in cell $(i,j)$.  (If any table has $X_{0,0}+X_{0,1} = 0$ or $= s$, we discard the table and draw a new one, because that would imply that one of the rows of the table is empty, in which case the Pearson Chi-Squared test statistic is not defined.  Similarly, if any table has $X_{0,0}+X_{1,0} = 0$ or $=s$, we discard the table and draw a new one.)
For each of the tables in the resulting sample, a Pearson chi-square test statistic is calculated as $T = \sum_{i=0}^1\sum_{j=0}^1 \frac{(X_{i,j}-q_{i,j})^2}{q_{i,j}}$, where $X_{i,j}$ is the observed count in cell $(i,j)$.  For each scenario, this results in $n=1000$ test statistics, $T_1, \ldots, T_n$, one for each table.  From these, we obtain $n=1000$ $p$~values, $p_1, \ldots, p_n$ by applying the $\chi^2_1$ approximation, i.e., $p_i = 1-F(T_i)$ for $i=1,\ldots, n$, where $F$ is taken to be the cdf of the $\chi^2_1$ distribution.\\

Figure \ref{fig:ex2stdaxes} shows the resulting Q-Q plots for scenarios 1 (in blue) and 2 (in red), where the 45$^{\mbox{\footnotesize o}}$ line is shown as well as the testing band obtained from the equal local levels procedure for testing, at global level .05, the null hypothesis that $p_1, \ldots, p_n$ have the same distribution as $n$ i.i.d.\ draws from U(0,1).  In Figure \ref{fig:ex2stdaxes}, the Q-Q plot for scenario 2 is made first,
and then the Q-Q plot for scenario 1 is added to the same axes by setting the \code{add}
argument of \pkg{qqconf} to \code{TRUE}, as follows:

\begin{figure}
\begin{center}
\includegraphics[width=.67\linewidth]{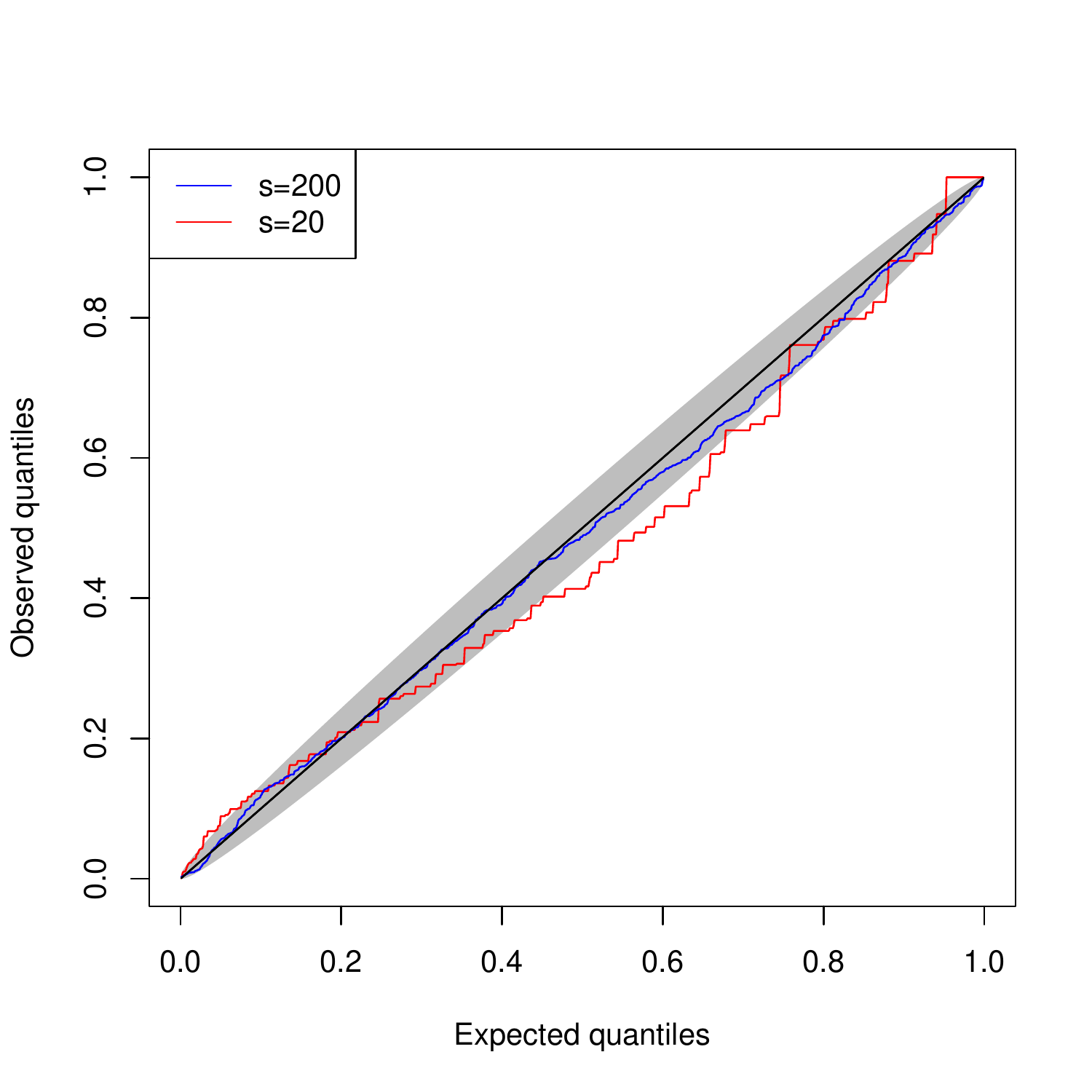}
\end{center}
\caption{Q-Q plots for scenarios 1 ($s =$ 200) and 2 ($s =$ 20) with level .05 testing band and standard axes.}
\label{fig:ex2stdaxes}
\end{figure}

\begin{Schunk}
\begin{Sinput}
R> pvals_scenario_1 <- scan("data/pvals_scenario_1", quiet = TRUE)
R> pvals_scenario_2 <- scan("data/pvals_scenario_2", quiet = TRUE)
R> qqconf::qq_conf_plot(
+    obs = pvals_scenario_2,
+    distribution = qunif,
+    points_params = list(col = "red", type="l")
+  )
R> qqconf::qq_conf_plot(
+    obs = pvals_scenario_1,
+    distribution = qunif,
+    points_params = list(col = "blue",type="l"),
+    add = TRUE
+  )
R> legend(
+    "topleft",
+    legend = c("s=200","s=20"),
+    col = c("blue","red"),
+    lty = 1
+  )
\end{Sinput}
\end{Schunk}

When assessing $p$~values, typically the lower tail is of most interest, but this part of the plot
is difficult to see when the plot axes are on the original scale. To focus the visualization on
the small $p$~values we
can plot the axes on the -log10 scale, as in Figure \ref{fig:ex2logaxes},
by setting the \code{log10} argument of \code{qqconf} to \code{TRUE}.
(Note that in Figure \ref{fig:ex2logaxes}, small $p$~values are to the top and right of the plot, so a curve that is too low is conservative, and too high is anti-conservative.)
Figure \ref{fig:ex2logaxes} is generated as follows:

\begin{figure}
\begin{center}
\includegraphics[width=.67\linewidth]{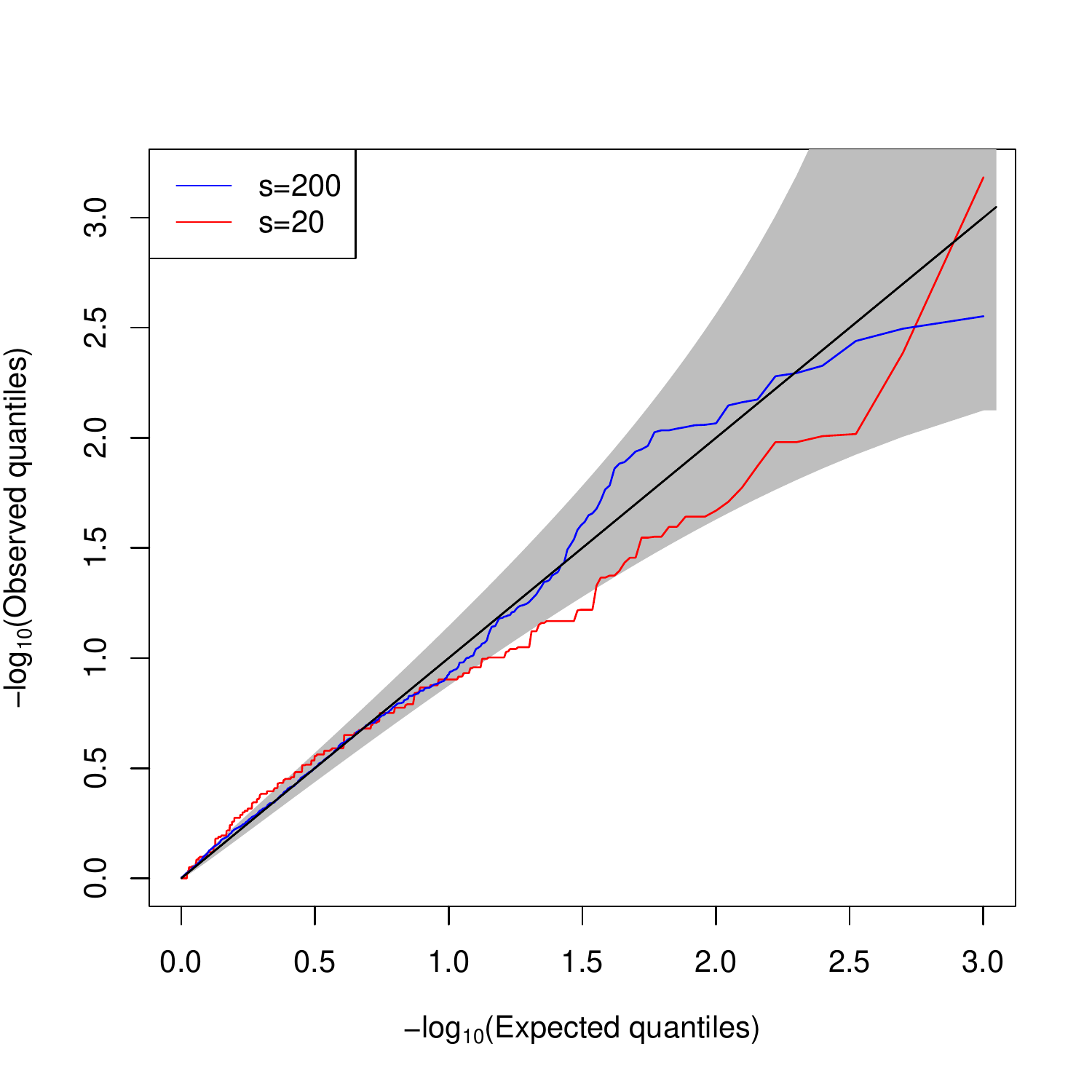}
\end{center}
\caption{Q-Q plots for scenarios 1 ($s =$ 200) and 2 ($s =$ 20) with axes on the $-\log_{10}$ scale.}
\label{fig:ex2logaxes}
\end{figure}

\begin{Schunk}
\begin{Sinput}
R> qqconf::qq_conf_plot(
+    obs = pvals_scenario_2,
+    distribution = qunif,
+    points_params = list(col = "red", type = "l"),
+    log10 = TRUE
+  )
R> qqconf::qq_conf_plot(
+    obs = pvals_scenario_1,
+    distribution = qunif,
+    points_params = list(col = "blue", type = "l"),
+    log10 = TRUE,
+    add = TRUE
+  )
R> legend(
+    "topleft",
+    legend = c("s=200","s=20"),
+    col = c("blue","red"),
+    lty = 1
+  )
\end{Sinput}
\end{Schunk}

From Figures \ref{fig:ex2stdaxes} and \ref{fig:ex2logaxes}, it can be seen that in scenario 1, when $s=200$ and the smallest expected cell count is 12, there is no significant deviation of the $p$~values from i.i.d.\ U(0,1) under the null hypothesis.  In contrast, in scenario 2, when $s=20$ and the smallest expected cell count is 1.2, the $\chi^2_1$ asymptotic distribution is not an accurate approximation to the sampling distribution of $T$.  As a result, we can see in Figures \ref{fig:ex2stdaxes} and \ref{fig:ex2logaxes} that the $p$~values differ significantly from i.i.d.\ U(0,1) under the null hypothesis, with small $p$~values tending to be overly conservative, while the larger $p$~values tend to be anti-conservative.

\subsection{Q-Q plots for $p$ values from genome-wide association studies}
\label{sec:gwas}
The goal of a genome-wide association study (GWAS) is to identify genetic variants
that influence a trait (where a trait is commonly a disease or some other measured variable such as blood pressure or blood glucose level). For each individual in the sample,
trait data are collected as well as genetic data on a large number of single-nucleotide
polymorphisms (SNPs) throughout the genome.  Based on these
data,
a statistical test is typically performed for
each SNP to assess whether it is associated with the trait, resulting in a large number of
$p$~values, one for each SNP.  Then a very stringent multiple testing correction is applied in order to declare a result for a SNP to be significant.  As part of the data analysis,
a Q-Q plot is commonly presented to visualize and assess the distribution of genome-wide
$p$~values.\\

The implicit null hypothesis being assessed in such a Q-Q plot is $H_0$: none of the tested
SNPs is associated with the trait.  If the SNPs could be assumed to be independent, then this
null hypothesis would correspond to the $p$~values being i.i.d.\ U(0,1) random variables.  One
could argue for either a one-sided or two-sided alternative.  The one-sided alternative would
be that there is an excess of small $p$~values (which is equivalent to $F(x) > x$ for some $x \in
(0,1)$ in the notation of Section \hyperref[sec:one-sided]{2.5}),
which would be biologically interpretable as indicating that at least one
SNP was associated with the trait.  The two-sided alternative would simply
be that the distribution is non-uniform.  While an excess of large $p$~values would not have
any particular biological interpretation, it could indicate a problem with the data analysis,
e.g., use of an inappropriate statistical test or the unexpected failure of assumptions
underlying the test used.
\newline
\newline
In fact, there is local correlation among genome-wide
SNPs, which decays very rapidly with distance as a result of genetic recombination.  Typically, some
``pruning''
is done on the genome-wide SNPs prior to analysis so that the remaining SNPs in each small, local
region are less correlated.
While the remaining SNPs have a local correlation structure, this has only a weak effect at a
genome-wide scale, and a Q-Q plot with appropriate simultaneous
bounds can still provide a valuable
visualization tool to assess the extent and type of deviation from the null.  When substantial
correlation remains, an alternative is to create the testing band based on an ``effective number'' of independent SNPs, \code{neff}.  This could be done by setting \code{n=neff} in
\code{get_qq_band}.  (In this case, different x-coordinates would obviously need to be used for
plotting the bounds of the band than for plotting the points.)\\

In GWAS, if the $p$~values deviate from the null,
it can be very useful to view graphically how they deviate. For instance, if a few tests
yield unusually small $p$~values but the $p$~values from the bulk of the tests look
relatively uniform, this suggests that the genetic variation affecting the trait of
interest is likely driven by a relatively small number of genetic
variants. If, however, there are
some small deviations from uniformity throughout the $p$~value distribution, this could
indicate that the trait of interest is affected by a large number of genetic variants
that all play some small part in a complex biological process, or it could potentially
indicate that there are some
confounding variables that are not controlled for.
\newline
\newline
Either of the above two scenarios, representing very different alternative distributions,
could commonly arise in a GWAS, so
the ELL method is a desirable choice for a putting testing bands on the Q-Q plot
because it is
agnostic to the choice of alternative distribution. Moreover, since small $p$~values are often of great interest in GWAS because they
can indicate the genetic variants that have the greatest influence on the trait, the use of
ELL testing bands is far superior to use of KS testing bands because of the comparatively
greater tail sensitivity of the former.
\subsubsection{Application of equal local levels to Creutzfeld-Jakob Disease}
We downloaded the $p$~values from a GWAS of Creutzfeld-Jakob disease (CJD) in a sample of 4,110 cases and 13,569 controls \citep{jones2020identification}. Tests of association between risk
for the disease and genetic variants were done at 6,314,492 SNPs. Major genetic
risk loci were found on chromosomes 1 and 20. Here, we remove those chromosomes
from the results in order to be able focus on parts of the genome where we remain
uncertain about to what extent risk variants are present.
For convenience, we subsample the remaining SNPs to 10,000 approximately evenly spaced SNPs, which also helps ensure that correlation between SNPs is minimized.
\begin{Schunk}
\begin{Sinput}
R> cjd_df <- read.table("data/cjd_sample.txt", header = TRUE)
\end{Sinput}
\end{Schunk}

\begin{figure}
\begin{center}
\includegraphics[width=.67\linewidth]{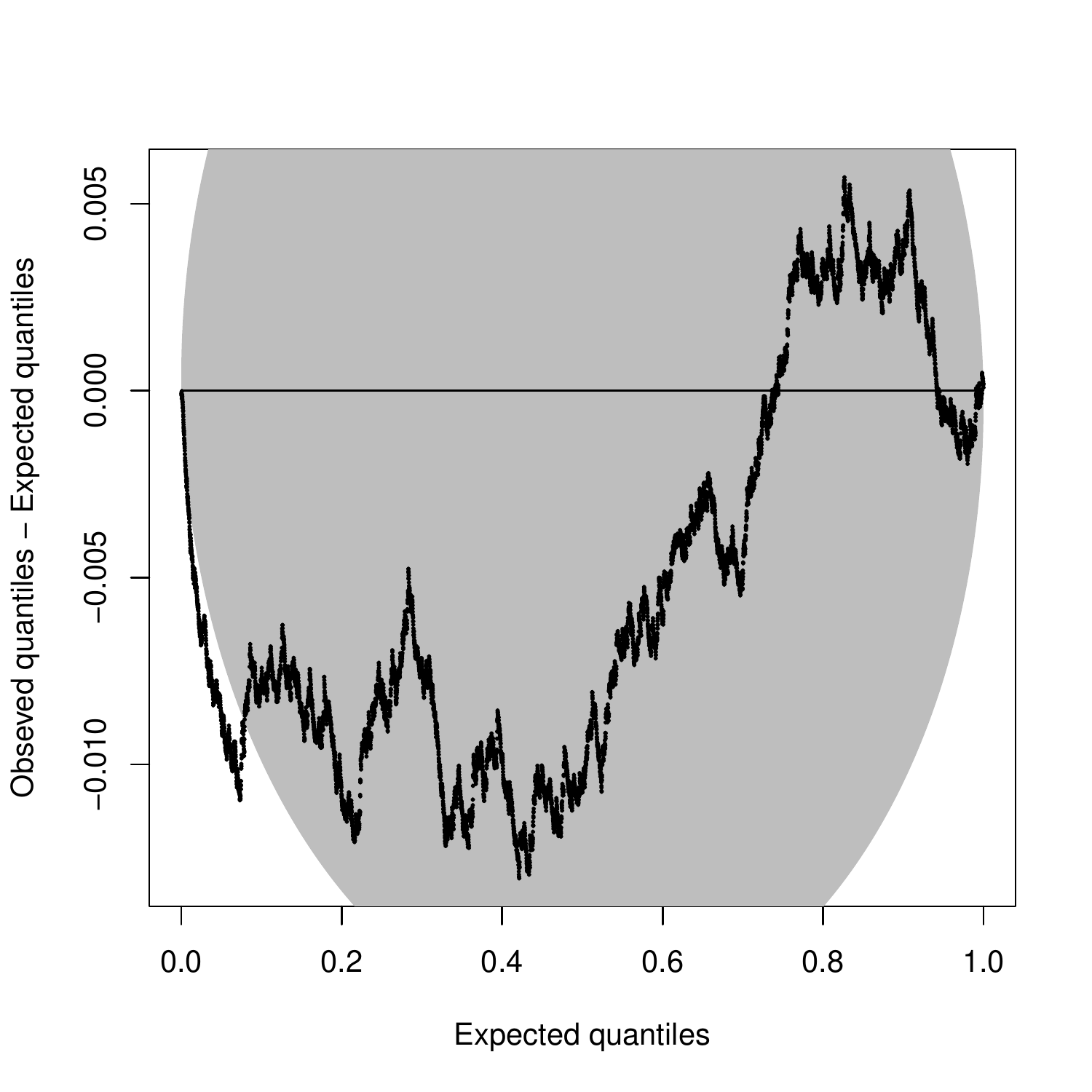}
\end{center}
\caption{Differenced Q-Q plot of CJD GWAS $p$~values.}
\label{fig:ex3ominuseaxes}
\end{figure}

We then make Q-Q plots of these 10,000 $p$~values with .05-level testing bands.
Note that for large datasets, a Q-Q plot with standard axes is undesirable,
because, e.g., the .05-level testing band becomes extremely close to the diagonal as $n$ grows,
so generally all the interesting information in the plot is
more-or-less collapsed onto the diagonal, rendering it
less effective as a visual tool.
For better visualization in a large dataset, we recommend
instead plotting the difference between the observed and expected quantiles versus the
expected quantiles, which we call a ``differenced'' Q-Q plot.
Such a plot can easily be created by setting the \code{difference} argument to
\code{TRUE} in \code{qq_conf_plot}.  Figure \ref{fig:ex3ominuseaxes}, which depicts the differenced Q-Q
plot for the CJD data, is produced as follows:
\begin{Schunk}
\begin{Sinput}
R> qqconf::qq_conf_plot(
+    obs = cjd_df[,3],
+    distribution = qunif,
+    points_params = list(pch = 21, cex = 0.2),
+    difference = TRUE
+    )
\end{Sinput}
\end{Schunk}

\begin{figure}
\begin{center}
\includegraphics[width=.67\linewidth]{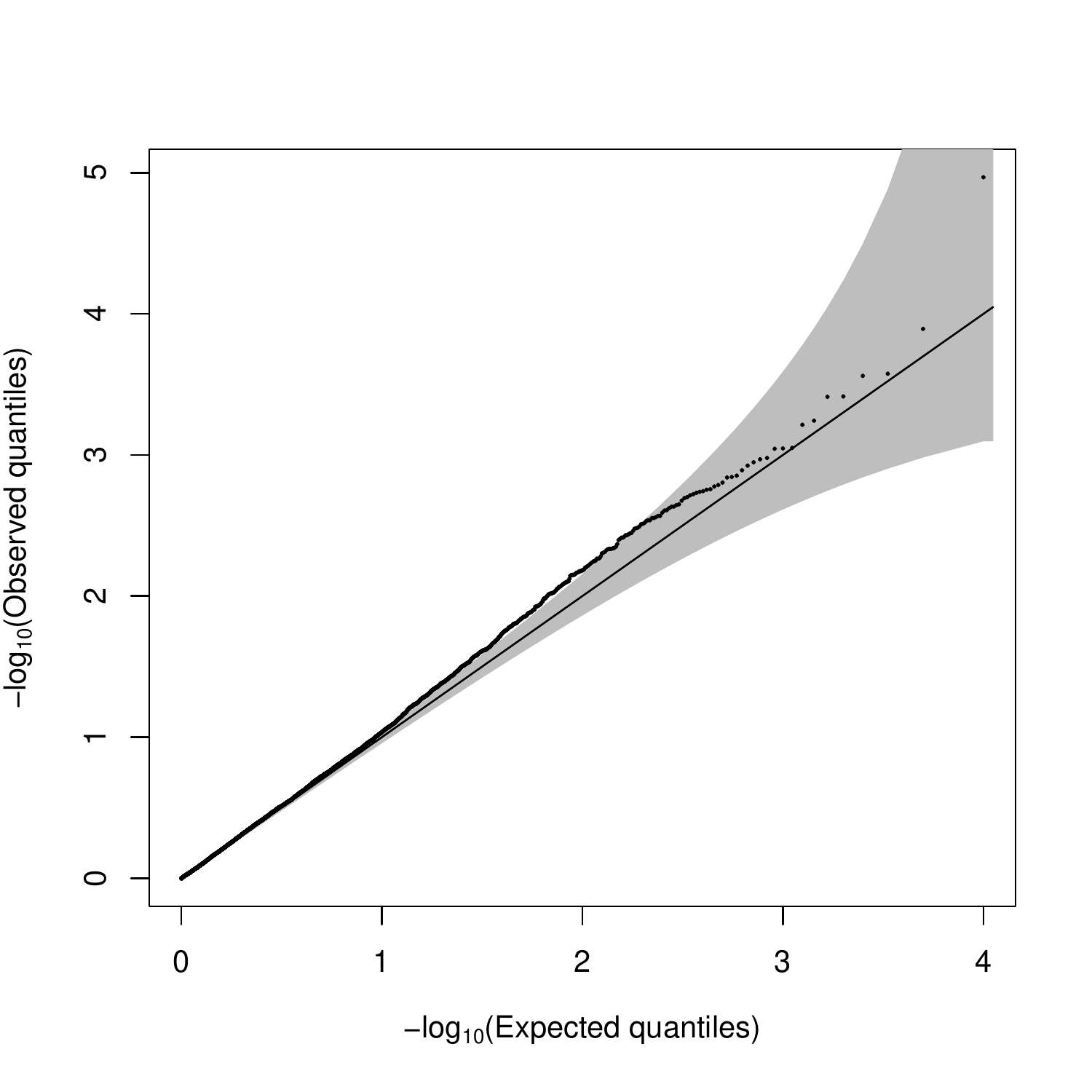}
\end{center}
\caption{Q-Q plot of CJD GWAS $p$~values, with axes on the -$\textrm{log}_{10}$ scale.}
\label{fig:ex3logaxes}
\end{figure}

\begin{figure}
\begin{center}
\includegraphics[width=.67\linewidth]{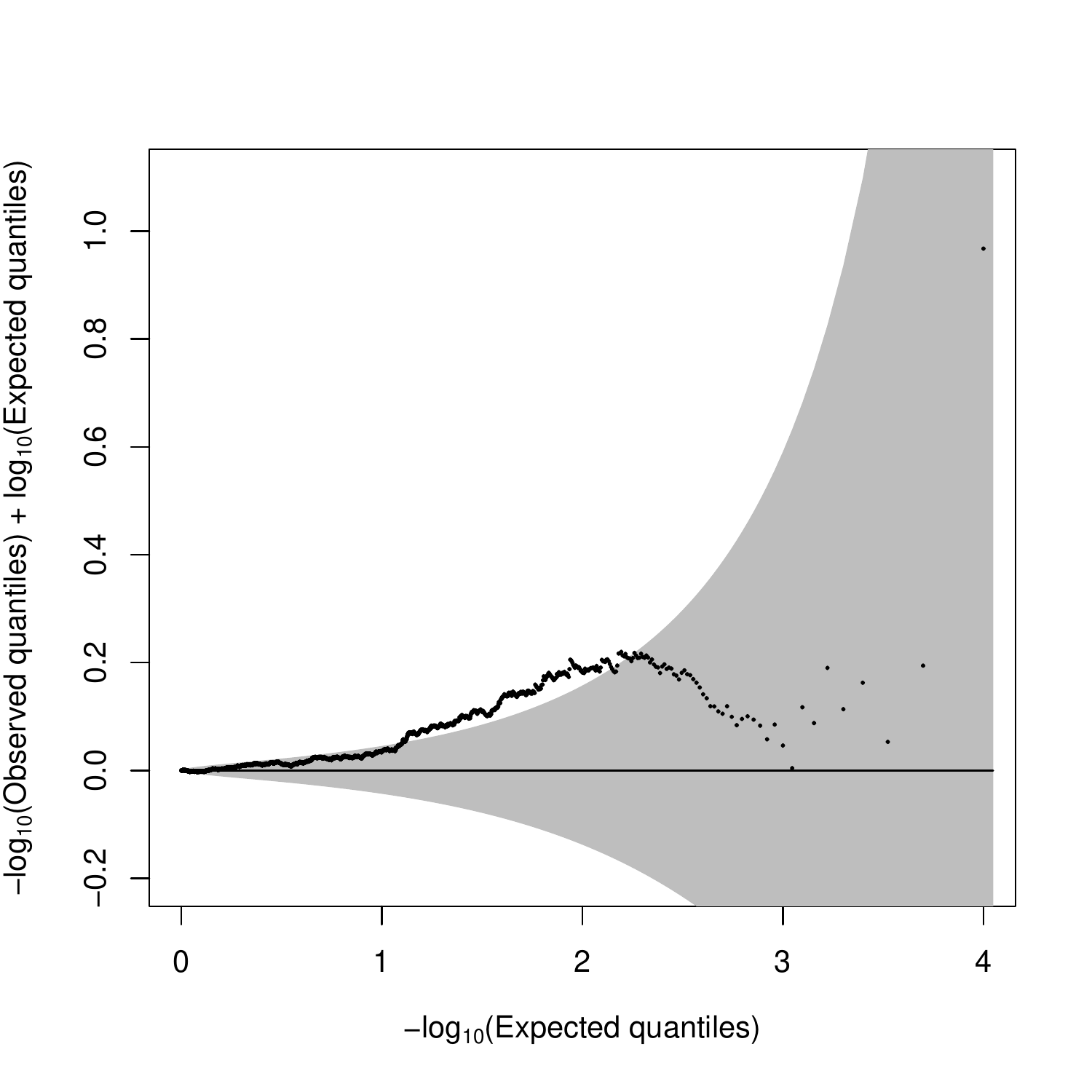}
\end{center}
\caption{Differenced Q-Q plot of CJD $p$~values, with quantiles on the -$\textrm{log}_{10}$ scale.}
\label{fig:ex3ominuselogaxes}
\end{figure}

In a GWAS the lower tail of the $p$~value distribution would typically represent
the most important genetic variants, so to
highlight this region, we
can use the \code{log10} argument to plot the axes on the -log10 scale for either a standard Q-Q plot (Figure \ref{fig:ex3logaxes})
or for a
differenced Q-Q plot (Figure \ref{fig:ex3ominuselogaxes}).
To create
Figure \ref{fig:ex3logaxes} we use

\begin{Schunk}
\begin{Sinput}
R> qqconf::qq_conf_plot(
+    obs = cjd_df[,3],
+    distribution = qunif,
+    points_params = list(pch = 21, cex = 0.2),
+    log10 = TRUE
+    )
\end{Sinput}
\end{Schunk}

and to create Figure \ref{fig:ex3ominuselogaxes} we use

\begin{Schunk}
\begin{Sinput}
R> qqconf::qq_conf_plot(
+    obs = cjd_df[,3],
+    distribution = qunif,
+    points_params = list(pch = 21, cex = 0.2),
+    difference = TRUE,
+    log10 = TRUE,
+    ylim = c(-0.2, 1.1)
+    )
\end{Sinput}
\end{Schunk}

From the Q-Q plots, we can see that there is an excess of moderately small $p$~values,
indicating that the test statistics do not follow the null distribution.
The type of deviation observed is
suggestive of a large number of sub-significant signals, likely representing genetic
variants that each contribute a small amount to the trait.  It is a common phenomenon
in GWAS of complex traits to have many small-effect SNPs whose signals
do not become significant except in very large sample sizes.

\section{Discussion}
\label{sec:discussion}
A Q-Q plot can be extremely valuable as a visualization tool for understanding
the extent and
type of deviation of a data set from a given reference distribution.
A crucial part of the interpretation of a Q-Q plot is the ability to distinguish run-of-the-mill
sampling variability from meaningful deviation, and this can be
accomplished by adding an appropriate testing band to a Q-Q plot.
ELL testing bands have been shown to be a notable improvement over
other available methods such as KS, but previously available software has been limited
to the normal distribution and is somewhat slow because it uses simulation to create
the bands.
To address the need for rapid generation of testing bands for Q-Q plots for a variety of
reference distributions, we have developed \pkg{qqconf}, an \proglang{R} package for creating
Q-Q plots, which is available on CRAN.
A notable feature of \pkg{qqconf} is the option to quickly and easily add a simultaneous
testing band based on ELL to
a Q-Q or P-P plot, for any reference distribution with a quantile function
implemented.  We show how \code{qqconf} can easily be used to output bands for use in other
plotting packages.  For the most common testing levels of .05 or .01, generation of testing
bands with \code{get_qq_band} in
\pkg{qqconf} is so fast that one can confidently generate such bands as a
default when creating Q-Q plots.\\

\pkg{qqconf} makes various accommodations for large data sets,
including (1) use of pre-computed and/or asymptotic values for
even faster implementation of testing bands in the function \code{get_qq_band}; 
and (2) the option to easily display Q-Q and
P-P plots on the difference scale for better visualization of large data sets.  For
applications in genomics (Section \hyperref[sec:gwas]{3.3}) and
assessing accuracy of $p$~values (Section \hyperref[sec:accuracy]{3.2}), one is particularly
interested in visualizing deviations in the tail of the distribution. In such cases, it is
particularly informative to view the Q-Q or P-P plot on a log scale (where the details of
the log transformation depend on which tail is of interest).
\pkg{qqconf} gives the option to easily generate such a log-transformed Q-Q or P-P plot to
focus on deviation in either the left or right tail.\\

Beyond the Q-Q plot application, ELL is a generic global testing method, and the problem
of determining, for a given number of local tests $n$ and a given global testing level $\alpha$,
the appropriate local level $\eta_n(\alpha)$ for an ELL test can arise in other applications,
particularly in genomics.  The \pkg{qqconf} package contains generic functions
(\code{get_bounds_two_sided}, \code{get_bounds_one_sided},
\code{get_level_from_bounds_two_sided} and
\code{get_level_from_bounds_one_sided})
to quickly
obtain $\eta_n(\alpha)$ for both one-sided and two-sided testing problems, using the FFT method \citep{moscovichnadler2016} as implemented in \cite{crossprob}.
For two-sided ELL in the cases of $\alpha =$ .05 and .01, we
have used the method of Appendix \hyperref[sec:appa]{A} to generate
extensive look-up tables for $n$ as large as 1 million and 500K, respectively,
and this permits quick access to these values or quick linear interpolation to approximate between
the grid points in cases where the grid is not saturated (e.g., near the largest values of $n$).
In addition, we have refined and applied previous
asymptotic approximations for two-sided ELL, which can
be confidently used for data sets of size 100K or larger.\\

In practice, we find that Q-Q plots are most often used either for distributions with known
parameters, such as U(0,1) or $\chi^2$ with known degrees of freedom, or for the normal distribution with
unknown parameters. \pkg{qqconf} provides extremely accurate testing bands for all such cases.
For the case of non-normal, non-uniform
reference distributions with unknown parameters, if the quantile/cdf/density
functions are implemented in \proglang{R}, then by default
\pkg{qqconf} will use maximum likelihood estimation
to estimate the parameters (though the user can easily substitute estimates of their choice)
and then form the testing band by taking these estimates as known
values.  In sufficiently large data sets, standard asymptotic theory ensures that the parameter
estimates will be close to
the true values, and this method will work well.  In small sample sizes, use of maximum likelihood
estimation in this context (non-normal distributions with unknown parameters)
tends to lead to overly conservative testing bands.  However, we have not yet
identified a substantive application that requires a non-normal reference distribution with
unknown parameters, so we do not know if this is of sufficient interest to warrant further
extensions.  If a need for this were identified, then two possible approaches to making the testing
bands less conservative for that situation would be (1) for
each distribution of interest, identify or develop a parameter estimation method that can be shown to
generate bands with the appropriate global level (as we have already done for the normal distribution)
or (2) extend the simulation-based approach of \cite{aldor2013power} to the distribution of interest,
where this could also involve choosing an appropriate estimation method as in (1).  A different
potential extension of arguably greater interest is to dependent rather than i.i.d.\ data, and for the
case of multivariate normal data where the
covariance structure is known or could be estimated, a simulation-based approach along the lines
of \cite{akinbiyi2020} could be
developed.

\section{Acknowledgments}
This study was supported by National Institutes of Health grant R01 HG001645 (to M.S.M.).

\bibliography{refs081022}

\section*{A. Recursion to compute global level of two-sided ELL}
\label{sec:appa}

First, note that because of property (\ref{wolog}),
without loss of generality, we can assume
\begin{equation}
    X_{1}, ..., X_{n} \distras{iid} U(0, 1).  \label{wolog2}
\end{equation}

The goal is to calculate the following probability:
\begin{equation*}
    \alpha = P_{0}\Big(\bigcup\limits_{i=1}^{n} \{X_{(i)} \notin (h_{i}, g_{i})\}\Big)
= 1 - P_{0}\Big(\bigcap\limits_{i=1}^{n} \{ X_{(i)} \in (h_{i}, g_{i})\}\Big) = 1 - P_{0}\Big(\bigcap\limits_{i=1}^{n} \{ X_{(i)} \in (h_{i}, g_{i}]\}\Big),
\end{equation*}
where $P_{0}$ is the probability distribution given by (\ref{wolog2}).
\newline

Let $b_{1}, ..., b_{2n}$ be the sorted values of $h_{1}, ..., h_{n}, g_{1}, ..., g_{n}$ in
ascending order. We also define $b_{0} = 0$ and $b_{2n + 1} = 1$. We divide the
interval $(b_{0}, b_{2n + 1})$ into $2n + 1$ bins, where bin 1 is $B_1 = (b_{0}, b_{1}]$,
bin 2 is $B_2 = (b_{1}, b_{2}], ...,$ and bin $2n + 1$ is $B_{2n+1} = (b_{2n}, b_{2n + 1})$.
Let $N_{j} = \sum_{i = 1}^{n}\mathds{1}(X_{i} \in B_j )$ denote the random variable that
counts the number of X's falling into bin j, for $1 \leq j \leq 2n+1$, and let $S_{k} =
\sum_{j=1}^{k} N_{j}$ be the $k$th partial sum of the $N$'s, for $1 \leq k \leq 2n+1$.
We make the following key observation:
\begin{equation*}
    \{X_{(i)} \in (h_{i}, g_{i}] \text{ for } i = 1, ..., n\} = \{l_{k} \leq S_{k} \leq u_{k} \textrm{ for } k = 1, 2, ..., 2n\},
\end{equation*}
where for $1 \leq k \leq 2n$,
\begin{align*}
u_{k} &=
    \begin{cases}
      0, & \text{if}\ k=1 \\
      \sum_{i = 1}^{k - 1} \mathds{1} \big(b_{i} \in \{h_{1}, ..., h_{n}\}\big), & \text{otherwise}
    \end{cases}\\
    l_{k} &=
      \sum_{i = 1}^{k} \mathds{1} \big(b_{i} \in \{g_{1}, ..., g_{n}\}\big)
\end{align*}
Note that $u_{2n} = l_{2n} = n$ always holds.  Here, $u_k$ is the number of order statistics whose
lower interval end points are to the left of bin $k$, so it is an upper bound on the number of
$X_i$'s that could occur in $\cup_{j=1}^k B_j$.  Similarly, $l_k$ is the number of order statistics
whose upper interval endpoints are to the left of bin $k+1$, so it is a lower bound on the number
of $X_i$'s that could occur in $\cup_{j=1}^k B_j$.
Thus, if we define $\Lambda = \{(m_1,\ldots, m_{2n}) \in \{0,\ldots, n\}^{2n}$ s.t.
$l_k
\leq s_k \leq u_k$ for $1 \leq k \leq 2n$, where $s_k = \sum_{i=1}^k m_i$ for $1 \leq k \leq 2n\}$, then
\[ P_0(\cap_{i=1}^n \{X_{(i)} \in (h_i,g_i]\}) = \sum_{(m_1,\ldots,m_{2n}) \in \Lambda}
P_0(N_j = m_j \mbox{ for } j=1,\ldots, 2n)\]
\[ = \sum_{(m_1,\ldots,m_{2n}) \in \Lambda} {n \choose m_1, \ldots, m_{2n}} \prod_{j=1}^{2n} (b_j -
b_{j-1})^{m_j},\]
a sum of probabilities of
multinomial events, where $m_j$ is the number of $X_i$'s that fall into bin $j$.
To calculate the needed probability, we define
\begin{equation*}
c_{j}^{(k)} = P_{0}(S_{k} = j \textrm{ and } l_{q} \leq S_{q} \leq u_{q} \textrm{ for } q = 1, ..., k - 1), \textrm{for } k = 1, ..., 2n \textrm{ and } j = 0, ..., n.
\end{equation*}
\[ \mbox{Then } c_{j}^{(k)} =
    P_{0}\Big(\bigcup\limits_{m=l_{k- 1}}^{min(j, u_{k- 1})} \{S_{k - 1} = m \textrm{ and } N_{k} = j - m \textrm{ and } l_{q} \leq S_{q} \leq u_{q} \textrm{ for } q = 1, ..., k - 2\}\Big)\]
\begin{flalign*}
    &= \sum\limits_{m=l_{k- 1}}^{min(j, u_{k- 1})}P_{0}(S_{k - 1} = m \textrm{ and } N_{k} = j - m \textrm{ and } l_{q} \leq S_{q} \leq u_{q} \textrm{ for } q = 1, ..., k - 2)\\
    &= \sum\limits_{m=l_{k- 1}}^{min(j, u_{k- 1})}c_{m}^{(k - 1)} * P_{0}(N_{k} = j - m | S_{k - 1} = m)\\
    &= \sum\limits_{m=l_{k- 1}}^{min(j, u_{k- 1})}c_{m}^{(k - 1)} * P(B = j - m), \textrm{where }B \sim \mbox{Binomial}\left(n - m, \frac{b_{k} - b_{k - 1}}{1 - b_{k - 1}}\right),
\end{flalign*}
which gives an easily computed recursive formula, where the initialization is
$c_{0}^{(1)} = (1 - b_{1})^{n}$.\\

In the case of general vectors $((h_{1}, ..., h_{n})$ and $(g_{1}, ..., g_{n}))$, subject only to
$0 \leq h_i < g_i \leq 1$ for $1 \leq i \leq n$, we could use the recursion to obtain $c_{n}^{(2n)}$,
and then obtain the global level $\alpha = 1 - c_{n}^{(2n)}$.  For the special case in which $(h_{1}, g_{1}),
\ldots, (h_{n}, g_{n})$ are derived from two-sided ELL,
i.e., Equations
(\ref{equallevels}) and (\ref{equalsides}) hold, then as a result of the symmetry in the problem,
for each $1 \leq j \leq n$, we
need only calculate $c_{j}^{(k)}$ for $k = 1, ..., n+1$ instead of $k = 1, \ldots , 2n$ and then use
\begin{equation*}
    1-\alpha = P_{0}\Big(\bigcap\limits_{i=1}^{n} X_{(i)} \in (h_{i}, g_{i})\Big) = \sum_{j = l_{n}}^{u_{n}} c_{j}^{(n)} \cdot \frac{c_{n - j}^{(n + 1)}}{\binom{n}{j}b_n^j(1 - b_{n})^{n-j}}.
\end{equation*}

To show this, we first define the following values for $k=1, \ldots 2n$:
\begin{align*}
\Tilde{u}_{k} &=
    \begin{cases}
      0, & \text{if}\ k = 2n \\
      \sum_{i = k + 1}^{2n} \mathds{1} \big(b_{i} \in \{g_{1}, ..., g_{n}\}\big), & \text{otherwise}
    \end{cases}\\
    \Tilde{l}_{k} &=
      \sum_{i = k}^{2n} \mathds{1} \big(b_{i} \in \{h_{1}, ..., h_{n}\}\big)\\
    T_{k} &= \sum_{j = k + 1}^{2n + 1}N_{j} = n - S_{k}
\end{align*}
Now, we make the following observations:
\newline
\newline
(1) With two-sided ELL, $g_{i} = 1 - h_{n + 1 - i}$ for $i = 1, ..., n$ because $F_{Beta(i, n + 1 - i)}^{-1}(1 - \frac{\eta}{2})$ = $1 - F_{Beta(n + 1 - i, i)}^{-1}(\frac{\eta}{2})$.
\newline
\newline
(2) $b_{k} = 1 - b_{2n + 1 - k}$ for $k = 1, ..., 2n$ by (1).
\newline
\newline
(3) $u_{k} = \Tilde{u}_{2n + 1 - k}$ and $l_{k} = \Tilde{l}_{2n + 1 - k}$, for $k = 1,..., 2n$ by (1) and (2).
\newline
\newline
(4) The random vector $(N_{1}, ..., N_{k})$ has the same distribution as $(N_{2n + 1}, ..., N_{2n+2-k})$ for $k = 1, ..., 2n + 1$. This follows from the fact that the vector $(X_{1}, ..., X_{n})$ has the same distribution as $(1 - X_{1}, ..., 1 - X_{n})$ (since each $X_{i}$ is independent uniform) and (2).
\newline
\newline
(5) The random vector $(S_{1}, ..., S_{k})$ has the same distribution as $(T_{2n}, ..., T_{2n + 1 - k})$ for $k = 1, ..., 2n$. This follows from (4).
\newline
\newline
(6) $c_{j}^{(k)}
    = P_{0}(T_{2n + 1 - k} = j \textrm{ and } \Tilde{l}_{r} \leq T_{r} \leq \Tilde{u}_{r} \textrm{ for } r = 2n + 2 - k, ..., 2n)$, which follows from (3) and (5).
\newline
\newline
(7) Conditional on $S_{k}$, the random vector $(X_{(S_{k} + 1)}, ..., X_{(n)})$ is distributed
as the order statistics of $n - S_{k}$ i.i.d.\ draws from $U(b_{k}, 1)$.
\newline
\newline
(8) The random vector $(S_{1}, ..., S_{r})$ and the random vector $(T_{r}, ..., T_{n})$ are independent conditional on $S_{r}$. This follows directly from (7).
\newline
\newline
Combining the above results, we can write
\begin{equation*}
    \{X_{(i)} \in (h_{i}, g_{i}) \textrm{ for } i = 1,..., n\} = \{l_{k} \leq S_{k} \leq u_{k} \textrm{ for } k = 1, ..., 2n\}.
\end{equation*}
Also, observe that for any $2 \leq r \leq 2n - 1$, we have
    $\{l_{k} \leq S_{k} \leq u_{k} \textrm{ for } k = 1, ..., 2n\} =$
\begin{equation*}
    \{l_{k} \leq S_{k} \leq u_{k} \textrm{ for } k = 1, ..., r \textrm{ and } T_{r} = n - S_{r} \textrm{ and } \Tilde{l}_{k} \leq T_{q} \leq \Tilde{u}_{k} \textrm{ for } q = r, ..., 2n\}.
\end{equation*}
Thus, we can write
$    \sum_{j = l_{r}}^{u_{r}} P(S_{r} = j \textrm{ and } l_{k} \leq S_{k} \leq u_{k} \textrm{ for } k = 1, ..., r - 1)$
    \[\cdot I(\Tilde{l}_{r} \leq n - j \leq \Tilde{u}_{r})
    \cdot P(\Tilde{l}_{q} \leq T_{q} \leq \Tilde{u}_{r} \textrm{ for } q = r + 1,..., 2n | T_{r} = n - j)\]
    \[= \sum_{j = l_{r}}^{u_{r}} c_{j}^{(r)} \cdot I(\Tilde{l}_{r} \leq n - j \leq \Tilde{u}_{r}) \cdot P(\Tilde{l}_{q} \leq T_{q} \leq \Tilde{u}_{r} \textrm{ for } q = r + 1,..., 2n | T_{r} = n - j) \]
    \[= \sum_{j = l_{r}}^{u_{r}} c_{j}^{(r)} \cdot I(\Tilde{l}_{r} \leq n - j \leq \Tilde{u}_{r}) \cdot \frac{P(\Tilde{l}_{q} \leq T_{q} \leq \Tilde{u}_{r} \textrm{ for } q = r + 1,..., 2n \textrm{ and } T_{r} = n - j)}{T_{r} = n - j}\]
    \[= \sum_{j = l_{r}}^{u_{r}} c_{j}^{(r)} \cdot \frac{c_{n - j}^{(2n + 1 - r)}}{\binom{n}{j}b_{r}^{j}(1 - b_{r})^{n-j}}\]
Now, if we let $r = n$ above, then we get
\begin{equation*}
    P_{0}\Big(\bigcap\limits_{i=1}^{n} X_{(i)} \in (h_{i}, g_{i})\Big) = \sum_{j = l_{n}}^{u_{n}} c_{j}^{(n)} \cdot \frac{c_{n - j}^{(n + 1)}}{\binom{n}{j}b_n^j(1 - b_{n})^{n-j}}
\end{equation*}
as claimed.

For the calculation of general boundary crossing probabilities, this type of
algorithm requires $O(n^3)$ operations.  However, for the ELL boundary crossing
problem, based on experiments involving a dense grid of
values of $n$ between $10$ and $50,000$ and $\alpha = .05$,
we find that the number of recursive steps required
is approximately $8n^{2}$. While each recursive step itself requires
calculating a binomial probability which is $O(n)$ due to the calculation of the
binomial coefficient multiplied by a quantity with powers as large as $n$,
these calculations can be memoized with $O(n^{2})$ cost. Thus, in our specific
context of ELL-based boundaries and for the range of sample sizes in our application,
we find that the computational time is approximately a constant multiple of $n^2$.
\newline

To find the value of $\eta$ for which the local level is $\alpha$, we perform a
binary search over the range ($\eta_{lower},\eta_{upper})$, where $\eta_{upper} = \alpha$ and
$\eta_{lower} = \frac{\alpha}{n}$, which is the lower bound given by the Bonferroni correction.
Under the assumption in Equation (\ref{wolog2}), note that $F_{0i}$ is Beta($i$,$n-i+1$), which is used
to obtain $h_i$ and $g_i$ via Equation (\ref{equalsides}).

\section*{B. Recursions to compute global level of one-sided ELL}
\label{sec:appb}

We describe an exact recursion to calculate
$\alpha^{'}_n(\eta)$ as well as an approximation, also recursive, which
is much faster and bounds the relative error in the reported global significance level to a tolerance set
by the user.
Again, because of the property in Equation (\ref{wolog}),
without loss of generality, we assume that under the null hypothesis,
$X_{1}, ..., X_{n} \distras{iid} U(0, 1). $
Given a proposed set of lower bounds $h_{1}, ..., h_{n}$, where $h_1 < \ldots < h_n$, the
goal is to
calculate the following probability
\begin{equation*}
    \alpha = P_{0}\Big(\bigcup\limits_{i=1}^{n} \{X_{(i)} < h_{i}\}\Big)
    = 1 - P_{0}\Big(\bigcap\limits_{i=1}^{n} \{ X_{(i)} \geq h_{i}\}\Big)
    = 1 - P_{0}\Big(\bigcap\limits_{i=1}^{n} \{ X_{(i)} > h_{i}\}\Big),
\end{equation*}
where $P_{0}$ is the probability under the null hypothesis $H_0:$ $X_1, \ldots X_n$ i.i.d.\ $U(0,1)$.\\

Similar to the two-sided case, we divide the interval $[0, 1]$ into $n+1$ bins. First, we define $h_{0} = 0$ and $h_{n + 1} = 1$. Now, suppose bin 1 is $B_1 = (h_{0}, h_{1}]$,
bin 2 is $B_2 = (h_{1}, h_{2}], ...,$ and bin $n + 1$ is $B_{n+1} = (h_{n}, h_{n + 1})$.
Let $N_{j} = \sum_{i = 1}^{n}\mathds{1}(X_{i} \in B_j )$ denote the random variable that
counts the number of X's falling into bin j, for $1 \leq j \leq n+1$, and let $S_{k} =
\sum_{j=1}^{k} N_{j}$ be the $k$th partial sum of the $N$'s, for $1 \leq k \leq n+1$. Similar to the two-sided case, we observe that the following two events are the same:
\begin{equation*}
    \{X_{(i)} > h_{i} \text{ for } i = 1, ..., n\} = \{S_{k} \leq k - 1 \textrm{ for } k = 1, 2, ..., n\},
\end{equation*}
Thus, if we define $\Lambda = \{(m_1,\ldots, m_{n}) \in \{0,\ldots, n\}^{n}$ s.t. $w_{n}=n$ and $w_k \leq k$ for $1 \leq k \leq n$, where $w_k = \sum_{i=1}^k m_i$ for $1 \leq k \leq n\}$, then
\[ P_0(\cap_{i=1}^n \{X_{(i)} > h_{i}\}) = \sum_{(m_1,\ldots,m_{n}) \in \Lambda}
P_0(N_1= 0 \mbox{ and } N_j = m_{j-1} \mbox{ for } j=2,\ldots, n+1)\]
\[ = \sum_{(m_1,\ldots,m_{n}) \in \Lambda} {n \choose m_1, \ldots, m_{n}} \prod_{j=1}^{n} (h_{j+1} -
h_j)^{m_j},\]
a sum of probabilities of
multinomial events, where $m_j$ is the number of $X_i$'s that fall into bin $j+1$.

\subsection*{B.1. Recursion for exact calculation}

For exact calculation of the needed probability, we define, for $k= 1,\ldots,n+1$,
\begin{equation*}
    c_{j}^{(k)} = P_{0}(S_{k} = j \textrm{ and } S_{l} \leq l - 1 \textrm{ for } l = 1, ..., k - 1),
\mbox{ for }j = 0, ..., k-1, \mbox{ and } c_k^{(k)} = 0.
\end{equation*}
Then for $0 \leq j \leq k-1$,
\begin{flalign*}
    \indent\indent c_{j}^{(k)} &= P_{0}(S_{k} = j \textrm{ and } S_{q} \leq q - 1 \textrm{ for } q = 1, ..., k - 1)\\
    &= P_{0}\Big(\bigcup\limits_{m=0}^{j} \{S_{k - 1} = m \textrm{ and } N_{k} = j - m \textrm{ and } S_{q} \leq q - 1 \textrm{ for } q = 1, ..., k - 2\}\Big)\\
    &= \sum\limits_{m=0}^{j}P_{0}(S_{k - 1} = m \textrm{ and } N_{k} = j - m \textrm{ and } S_{q} \leq q - 1 \textrm{ for } q = 1, ..., k - 2)\\
    &= \sum\limits_{m=0}^{j}c_{m}^{(k - 1)} * P_{0}(N_{k} = j - m | S_{k - 1} = m)\\
    &= \sum\limits_{m=0}^{j}c_{m}^{(k - 1)} * P(B = j - m), \textrm{where }B \sim \mbox{Binomial}\left(n - m, \frac{h_{k} - h_{k - 1}}{1 - h_{k - 1}}\right),
    &&
\end{flalign*}
which gives an easily computed recursive formula, where the initialization is ${c}_0^{(1)} = (1-h_1)^n$.
Then the global level $\alpha$ is equal to $1 - c_{n}^{(n+1)}$.

\subsection*{B.2. Recursion for fast approximation with error control}

For sufficiently large $k$, the terms of the sum for small $i$ in the update step
\[ c_j^{(k)} = \sum_{i=0}^j c_j^{(k-1)}* P(B= j-i), \mbox{ where } B \sim \mbox{Bin}\Big(n-i,
\frac{h_k - h_{k-1}}{1-h_{k-1}}\Big),\]
become negligible, and as $k$ gets large, we speed up the algorithm by dropping negligible terms while
bounding the relative error in the final calculation of the global level.
Because all terms in the sum are
positive, the approximation will be less than or equal to the true global level, and below we define the error
to be the global level minus the approximation, which will always be nonnegative.  This
will lead to a slightly conservative ELL test, but with the relative error in the level
guaranteed to be bounded by an arbitrary pre-specified amount.\\

As $k$ increases, we specify a
schedule for checking whether there are sufficiently small terms that can be dropped.  We begin checking at
$k = $ \code{first\_check}, and after that, we check whenever $k$ is a multiple of \code{check\_interval}.  At a given
checkpoint, the decision of whether additional terms can be dropped is based only on the current values of
the recursive variables $c^{(k)}_i$, $1 \leq i \leq k-1$ and the current value of \code{accumul\_err\_upper\_bnd},
which is an upper bound on the error in the global level due to the
terms that have already been dropped.  The checkpoints at which additional terms of the
sum are chosen to be dropped are termed ``drop points,'' and we label these $d_1, \ldots, d_w$, where
\code{first\_check} $\leq d_1 < \ldots < d_w < n$.
We end up with a modified recursion with initialization $\tilde{c}_0^{(1)} = (1-h_1)^n$ and
$\tilde{c}_1^{(1)} = 0$, and with update step:
\[ \tilde{c}_j^{(k)} = \sum_{i=\mbox{skip}_k+1}^j \tilde{c}_i^{(k-1)} * P(B = j-i),
\mbox{ where $B$ is Binomial($n-i$, $\frac{h_k-h_{k-1}}{1-h_{k-1}}$)},\]
for skip$_k+1 \leq j \leq k-1$
and $\tilde{c}_k^{(k)}=0$,
where skip$_k = -1$ for $1 \leq k \leq d_1$, skip$_k =T_i$ for $d_i < k \leq d_{i+1}$ and $1 \leq i \leq w-1$, and
skip$_k = T_w$ for $d_w < k \leq n$. In other words, at drop point $d_1$,
terms of the sum indexed 0 through $T_1$ are dropped,
and for $2 \leq m \leq w$, at drop point $d_m$ additional terms indexed by $T_{m-1}+1$ through $T_m$ are dropped
(terms 0 through $T_{m-1}$ having already been dropped at previous drop points).
At drop point $d_1$, the value $T_1$ is chosen to
be the largest value of $T$ such that
\begin{equation}
T < d_1 \mbox{ and } \frac{\sum_{j=0}^T \tilde{c}_j^{(d_1)}}{1 - \sum_{l=0}^{d_1-1} \tilde{c}_l^{(d_1)}} \leq \mbox{\code{max\_rel\_err}},\label{drop1}
\end{equation}
and for $2 \leq m \leq w$, at drop point $d_m$, the value $T_m$ is chosen to be the
largest value of $T$ such that
\begin{equation} d_m > T > T_{m-1}
\mbox{ and }
\frac{e_{m-1} + \sum_{j=T_{m-1}+1}^T \tilde{c}_j^{(d_m)}}{1-e_{m-1} - \sum_{l=T_{m-1}+1}^{d_m-1}
\tilde{c}_l^{(d_m)}} \leq \mbox{\code{max\_rel\_err}},\label{dropm}
\end{equation}
where we define $e_m$ to be the value of
\code{accumul\_err\_upper\_bnd} after recursion step $k = d_{m}$, which is given by
\[e_{1} = \sum_{j=0}^{T_{1}} \tilde{c}_j^{(d_{1})} \mbox{ and for } m>1, \,\, e_{m} = \sum_{j=0}^{T_{1}} \tilde{c}_j^{(d_{1})} + \sum_{i=2}^m \,\, \sum_{j=T_{i-1}+1}^{T_{i}}
\tilde{c}_j^{(d_{i})}.\]
A given checkpoint $k$ becomes a drop point if and only if there is some $T$
satisfying the corresponding constraints (either equation \ref{drop1} or \ref{dropm}).

We first show that $e_m$ is an upper bound on
the actual accumulated error $a_{m}$
that will be incurred in the calculation due to all terms dropped prior to drop point
$d_{m+1}$.  We define $E$ to be the event $\{S_k \leq k-1 \mbox{ for } k = 1,2, \ldots, n\}$.  Then, as noted in
Section \hyperref[sec:methods]{2}, $P(E) = c_n^{(n+1)}$, and the global level is 1-$P(E$).\\

{\bf Lemma:} $a_i \leq e_i$ for $i = 1, \ldots, w$.\\

{\bf Proof:}
Note that at drop point $d_1$, dropping terms of the sum indexed 0 through $T_1$ is equivalent to adding
an extra requirement that
$S_{d_1} > T_1$, so that instead of $P(E)$, we will be calculating $P(E \cap \{S_{d_1} > T_1\})$.
Therefore, the actual
accumulated error
in the global level
that will be incurred by this is $a_1 = P(E \cap \{S_{d_1} \leq T_1\})$, which is
bounded
above by $P(S_{d_1} \leq T_1 \mbox{ and } S_l \leq l-1 \mbox{ for } l=1,\ldots,d_1-1) = \sum_{j=0}^{T_1} c_j^{(d_1)} = \sum_{j=0}^{T_1} \tilde{c}_j^{(d_1)} = e_1$.\\

At the induction step, we assume that $a_{m-1} \leq e_{m-1}$.
Now consider the actual accumulated error $a_m$ due to the terms dropped at
drop points $d_1,\ldots, d_m$.  By similar logic as above,
\[a_m = P(E \cap [\cap_{j=1}^m (S_{d_j} > T_j)]^c) = P(E \cap \{[\cap_{j=1}^{m-1} (S_{d_j} > T_j)]^c \cup
(S_{d_m} > T_m)^c\})\]
\[ = P(E \cap [\cap_{j=1}^{m-1} (S_{d_j} > T_j)]^c) + P(E \cap [\cap_{j=1}^{m-1} (S_{d_j} > T_j) ]
\cap (S_{d_m} > T_m)^c)\]
\[ = a_{m-1} + P(E \cap [\cap_{j=1}^{m-1} (S_{d_j} > T_j) ]
\cap (S_{d_m} \leq T_m)) \leq a_{m-1} + \sum_{j=T_{m-1}+1}^{T_m} \tilde{c}_j^{(d_m)}\]
\[ \leq e_{m-1} + \sum_{j=T_{m-1}+1}^{T_m} \tilde{c}_j^{(d_m)} = e_m,\]

where the 2nd and 3rd equalities are based only on elementary set theory. $\square$\\

We now prove that this algorithm
guarantees relative error of no more than \code{max\_rel\_err} in the calculated global level.
First we note that
$P(E) = c_n^{(n+1)}$ satisfies
\[ c_n^{(n+1)} \leq \sum_{j=0}^{k-1} c_j^{(k)}, \mbox{ for all } 1 \leq k \leq n+1,\]
where this useful inequality follows directly from the definition of $c_j^{(k)}$ in Section
\hyperref[sec:methods]{2}.  A consequence is that $ P(E) \leq \sum_{j=0}^{d_1-1} c_j^{(d_1)}
 = \sum_{j=0}^{d_1-1} \tilde{c}_j^{(d_1)}$. Second, we note that by similar reasoning, for $m \geq 2$,
\[ P(E) = P(E \cap [\cap_{j=1}^{m-1} (S_{d_j} > T_j)]^c) + P(E \cap [\cap_{j=1}^{m-1} (S_{d_j} > T_j)])\]
\[ = a_{m-1} + P(E \cap [\cap_{j=1}^{m-1} (S_{d_j} > T_j)])
\leq a_{m-1} + \sum_{j=T_{m-1}+1}^{d_m-1} \tilde{c}_j^{(d_m)} \leq e_{m-1} + \sum_{j=T_{m-1}+1}^{d_m-1} \tilde{c}_j^{(d_m)}.\]

Therefore, the actual relative error in the global level incurred at drop point
$d_1$, which is $a_1/(1-P(E))$, satisfies
\[ \frac{a_1}{1-P(E)} \leq \frac{e_1}{1-\sum_{j=0}^{d_1-1}
\tilde{c}_j^{(d_1)}} = \frac{\sum_{j=0}^{T_1}\tilde{c}_j^{(d_1)}}{1-\sum_{j=0}^{d_1-1}
\tilde{c}_j^{(d_1)}}
\leq \mbox{\code{max\_rel\_err}} \]
by equation \ref{drop1}.
Furthermore, the actual relative error in the global level
that will be incurred in the calculation due to all terms dropped at or before $d_m$ is
\[a_m/(1-P(E)) \leq \frac{e_m}{1-e_{m-1}-\sum_{l=T_{m-1}+1}^{d_m-1} \tilde{c}_j^{(d_m)}}
= \frac{e_{m-1} + \sum_{j=T_{m-1}+1}^{T_m} \tilde{c}_j^{(d_m)}}{1-e_{m-1}-
\sum_{l=T_{m-1}+1}^{d_m-1} \tilde{c}_j^{(d_m)}}\]
\[\leq \mbox{max\_rel\_err}\]
by equation \ref{dropm}.  $\square$
The resulting algorithm is given below.
\begin{algorithm}[H]
\caption{Calculate approximate global level $\alpha$ for one-sided ELL from proposed lower bounds using speedup.}
\textbf{Input:} Vector of lower bound values where these must be increasing $(h_{1}, ..., h_{n})$, first value of $k$ for which to check for skipping \code{first\_check}, interval for which to check for skipping \code{check\_interval}, maximum allowed relative error in global level calculation \code{max\_rel\_err}.
\newline
\textit{get\_level\_from\_bounds\_one\_sided}$((h_{1}, ..., h_{n}), \textrm{ \code{first\_check}}, \textrm{ \code{check\_interval}}, \textrm{\code{max\_rel\_err}})$
\begin{algorithmic}[1]
\STATE {$c_{0}^{(1)} \leftarrow (1 - h_{1})^{n}$}
\STATE{$c_1^{(1)} \leftarrow 0$}
\STATE $\textrm{\code{accumul\_err\_upper\_bnd}} \leftarrow 0$
\STATE $\textrm{\code{skip}} \leftarrow -1$
\FOR{$k = 2, ..., n$}
\FOR{$j = skip + 1, ..., k - 1$}
\STATE $c_{j}^{(k)} \leftarrow 0 $
\FOR{$m = skip + 1, ..., j$}
\STATE $c_{j}^{(k)} \leftarrow  c_{j}^{(k)} + c_{m}^{(k - 1)} * \mbox{dbinom}(x = j - m, \mbox{size} = n-m, \mbox{prob} = \frac{(h_{k} - h_{k - 1})}{(1 - h_{k - 1})})$
\ENDFOR
\ENDFOR
\STATE {$c_k^{(k)} \leftarrow 0$}
\IF{$(k > \textrm{\code{first\_check}} \textbf{ and } k \textrm{ \% } \textrm{\code{check\_interval}} == 0) \textbf{ or } k == \textrm{\code{first\_check}}$}
\STATE $\textrm{\code{available\_err}} \leftarrow \textrm{\code{max\_rel\_err}} - (1 + \textrm{\code{max\_rel\_err}}) \cdot \textrm{\code{accumul\_err\_upper\_bnd}}$
\STATE $\textrm{\code{calculated\_total\_prob}} \leftarrow 0$
\FOR{$j = skip + 1, ..., k-1$}
\STATE $\textrm{\code{calculated\_total\_prob}} \leftarrow \textrm{\code{calculated\_total\_prob}} + c_{j}^{(k)}$
\ENDFOR
\STATE $\textrm{\code{available\_err}} \leftarrow \textrm{\code{available\_err}} - \textrm{\code{max\_rel\_err}} \cdot \textrm{\code{calculated\_total\_prob}}$
\STATE $\textrm{\code{proposed\_err}} \leftarrow c_{skip+1}^{(k)}$
\STATE $\textrm{\code{proposed\_skip}} \leftarrow \textrm{skip} + 1$
\WHILE{$\textrm{\code{proposed\_err}} \leq \textrm{\code{available\_err}}$}
\STATE{$\textrm{\code{proposed\_skip}} \leftarrow \textrm{\code{proposed\_skip}} + 1$}
\STATE{$\textrm{\code{proposed\_err}} \leftarrow \textrm{\code{proposed\_err}} + c_{\textrm{proposed\_skip}}^{(k)}$}
\ENDWHILE
\STATE $\textrm{\code{accumul\_err\_upper\_bnd}} \leftarrow \textrm{\code{proposed\_err}} - c_{\textrm{proposed\_skip}}^{(k)}$
\STATE $\textrm{\code{skip}} \leftarrow \textrm{\code{proposed\_skip}} - 1$
\ENDIF
\ENDFOR
\STATE $c_{n}^{(n+1)} \leftarrow 0$
\FOR{$l = skip + 1, ..., n - 1$}
\STATE $c_{n}^{(n+1)} \leftarrow c_{n}^{(n+1)} + c_{l}^{(n)}$
\ENDFOR
\STATE $\alpha \leftarrow 1 - c_{n}^{(n+1)}$
\RETURN $\alpha$
\end{algorithmic}
\textbf{end}
\end{algorithm}

\end{document}